\documentclass[fleqn]{article}
\usepackage{mystyle,mymath,proof,latexsym,amssymb,stmaryrd}

\begin{document}
\sloppy \hbadness=10000 \vbadness=10000


\def\Univ{{\rm Univ}}
\def\Left{{\rm Left}}
\def\Right{{\rm Right}}
\def\PC{{\rm PC}}
\def\Fin{{\rm Fin}}
\def\Decide{{\rm Decide}}
\def\CLJIDomega{{{\bf CLJID}^\omega}}
\def\CLJIDHA{{{\bf CLJID}^\omega+{\bf HA}}}
\def\LJIDHA{{{\bf LJID}+{\bf HA}}}
\def\LJID{{\bf LJID}}
\def\OP{{\rm OP}}
\def\Div{{\rm Div}}
\def\RP{{\rm RP}}
\def\MS{{\rm MS}}
\def\ET{{\rm ET}}
\def\LiftedTree{{\rm LiftedTree}}
\def\First{{\rm first}}
\def\Er{{\rm Er}}
\def\ER{{\rm ER}}
\def\Insert{{\rm insert}}
\def\ET{{\rm ET}}
\def\SInd{{\rm SInd}}
\def\Trans{{\rm Trans}}
\def\HA{{\bf HA}}
\def\DM{{\rm DM}}
\def\Ext{{\rm ext}}
\def\KB{{\rm KB}}
\def\DT{{\rm DT}}
\def\DS{{\rm DS}}
\def\Monoseq{{\rm Monoseq}}
\def\Eq{{\rm seq}}
\def\Last{{\rm last}}


\def\Uor{\biguplus}
\def\CLKIDomega{{{\bf CLKID}^\omega}}
\def\CLKIDPA{{{\bf CLKID}^\omega+{\bf PA}}}
\def\LKIDPA{{{\bf LKID}+{\bf PA}}}
\def\PA{{\bf PA}}


\def\LKIDN{{\bf LKIDN}}
\def\CLKIDN{{\bf CLKIDN}}


\def\Choose{{\rm Choose}}
\def\Asym{{\rm Asym}}
\def\Peano{{\bf Peano}}
\def\Code#1{{\lceil{#1}\rceil}}
\def\Ineq{{\rm Ineq}}
\def\MonoPath{{\rm MonoPath}}
\def\BoundedPath{{\rm BoundedPath}}
\def\InfPath{{\rm InfPath}}
\def\InfMonoPath{{\rm InfMonoPath}}
\def\HomSeq{{\rm HomSeq}}
\def\InfHomSeq{{\rm InfHomSeq}}

\def\ErdosTree{{\rm ErdosTree}}
\def\KB{{\rm KB}}
\def\MonoList{{\rm MonoList}}
\def\Seq{{\rm Seq}}
\def\Infinite{{\rm Infinite}}
\def\S{{\bf S}}
\def\N{N}
\def\Ind{{\rm Ind}}
\def\LKID{{\bf LKID}}
\def\PA{{\bf PA}}
\def\LKIDExt{{\bf LKIDExt}}
\def\CLKID{{\bf CLKID^\omega}}


\def\Over#1#2{\deduce{#2}{#1}}
\def\List{{\rm List}}
\def\ListX{{\rm ListX}}
\def\ListO{{\rm ListO}}
\def\ListE{{\rm ListE}}
\def\Nl{{\rm nl}}


\def\SLLID{{\rm SLID}}
\def\SLID{{\rm SLID}}
\def\CSLID{{\rm CSLID}}
\def\CSLIDomega{{{\rm CSLID}\omega}}
\def\Eclass{{\rm Eclass}}
\def\Eq{{\rm Eq}}
\def\Deq{{\rm Deq}}
\def\Satom{{\rm Satom}}
\def\Cells{{\rm Cells}}
\def\Roots{{\rm Roots}}
\def\Elim{{\rm Elim}}
\def\Case{{\rm Case}}
\def\Unfold{{\rm Unfold}}
\def\Pred{{\rm Pred}}
\def\Start{{\rm Start}}
\def\Unsat{{\rm Unsat}}
\def\Split{{\rm Split}}
\def\Underscore{\underline{\phantom{x}}}
\def\Rootcell{{\rm rootcell}}
\def\Rootshape{{\rm rootshape}}
\def\Jointleaf{{\rm jointleaf}}
\def\Jointleafimplicit{{\rm jointleafimplicit}}
\def\Jointnode{{\rm jointnode}}
\def\Directjoint{{\rm directjoint}}


\def\Range{{\rm Range}}
\def\Xstart{X_{{\rm start}}}
\def\kstart{k_{{\rm start}}}
\def\Leaves{{\rm Leaves}}
\def\PureDist{{\rm PureDist}}
\def\Pure{{\rm Pure}}
\def\Identity{{\rm Identity}}
\def\Subst{{\rm Subst}}


\def\Connected{{\rm Connected}}
\def\Eststablished{{\rm Eststablished}}
\def\Valued{{\rm Valued}}


\def\SLRDbtw{{\rm SLRD}_{btw}}
\def\BTW{{{\rm BTW}}}
\def\Loc{{{\rm Loc}}}
\def\Ne{{\ne}}
\def\Nil{{{\rm nil}}}
\def\eqDef{=_{{\rm def}}}
\def\Inf#1{{\infty_{#1}}}
\def\Stores{{\rm Stores}}
\def\SVars{{{\rm SVars}}}
\def\Val{{{\rm Val}}}
\def\MSO{{{\rm MSO}}}
\def\Sep{{{\rm Sep}}}
\def\Septwo{{{\rm SLMI}}}
\def\SLMI{{{\rm SLMI}}}
\def\Sepinf{{\rm Sep}\infty}
\def\THeaps{{\rm THeaps}}
\def\Root{{\rm Root}}
\def\TG{{\rm TGraph}}


\def\Nil{{{\rm nil}}}
\def\eqDef{=_{{\rm def}}}
\def\Inf#1{{\infty_{#1}}}
\def\Stores{{\rm Stores}}
\def\SVars{{{\rm SVars}}}
\def\Val{{{\rm Val}}}
\def\MSO{{{\rm MSO}}}
\def\Sep{{{\rm sep}}}
\def\THeaps{{\rm THeaps}}
\def\Root{{\rm Root}}
\def\TG{{\rm TGraph}}

\def\Tilde{\widetilde}
\def\Bar{\overline}
\def\Lequiv{\Longleftrightarrow}
\def\Lto{\Longrightarrow}
\def\Lfrom{\Longleftarrow}
\def\Norm{{\rm Norm}}
\def\Noshare{{\rm Noshare}}
\def\Roots{{\rm Roots}}
\def\Forest{{\rm Forest}}
\def\Var{{\rm Var}}
\def\Range{{\rm Range}}
\def\Cell{{\rm Cell}}
\def\Tree{{\rm Tree}}
\def\Switch{{\rm Switch}}
\def\All{{\rm All}}
\def\To{\leadsto}
\def\tree{{\rm tree}}
\def\Paths{{\rm Paths}}
\def\Finite{{\rm Finite}}
\def\Const{{\rm Const}}
\def\Leaf{{\rm Leaf}}
\def\LeastElem{{\rm LeastElem}}
\def\LeastIndex{{\rm LeastIndex}}
\def\WSnS{{{\rm WSnS}}}
\def\Expand{{{\rm Expand}}}


\long\def\J#1{} 
\def\T#1{\hbox{\color{green}{$\clubsuit #1$}}}
\def\W#1{\hbox{\color{Orange}{$\spadesuit #1$}}}

\def\Node{{{\rm Node}}}
\def\LL{{{\rm LL}}}
\def\DSN{{{\rm DSN}}}
\def\DCL{{{\rm DCL}}}
\def\LS{{{\rm LS}}}
\def\Ls{{{\rm ls}}}

\def\FPV{{{\rm FPV}}}
\def\Lfp{{{\rm lfp}}}
\def\IsHeap{{{\rm IsHeap}}}

\def\Equiv{\quad \equiv\quad }
\def\Null{{{\rm null}}}
\def\Emp{{{\rm emp}}}
\def\If{{{\rm if\ }}}
\def\Then{{{\rm \ then\ }}}
\def\Else{{{\rm \ else\ }}}
\def\While{{{\rm while\ }}}
\def\Do{{{\rm \ do\ }}}
\def\Cons{{{\rm cons}}}
\def\Dispose{{{\rm dispose}}}
\def\Vars{{{\rm Vars}}}
\def\Locs{{{\rm Locs}}}
\def\States{{{\rm States}}}
\def\Heaps{{{\rm Heaps}}}
\def\FV{{{\rm FV}}}
\def\True{{{\rm true}}}
\def\False{{{\rm false}}}
\def\Dom{{{\rm Dom}}}
\def\Abort{{{\rm abort}}}
\def\New{{{\rm New}}}
\def\W{{{\rm W}}}
\def\Pair{{{\rm Pair}}}
\def\Lh{{{\rm Lh}}}
\def\lh{{{\rm lh}}}
\def\Elem{{{\rm Elem}}}
\def\EEval{{{\rm EEval}}}
\def\PEval{{{\rm PEval}}}
\def\HEval{{{\rm HEval}}}
\def\EVal{{{\rm Eval}}}
\def\Domain{{{\rm Domain}}}
\def\Exec{{{\rm Exec}}}
\def\Store{{{\rm Store}}}
\def\Heap{{{\rm Heap}}}
\def\Storecode{{{\rm Storecode}}}
\def\Heapcode{{{\rm Heapcode}}}
\def\Lesslh{{{\rm Lesslh}}}
\def\Addseq{{{\rm Addseq}}}
\def\Separate{{{\rm Separate}}}
\def\Result{{{\rm Result}}}
\def\Lookup{{{\rm Lookup}}}
\def\ChangeStore{{{\rm ChangeStore}}}
\def\ChangeHeap{{{\rm ChangeHeap}}}
\def\Wand{\mathbin{\hbox{\hbox{---}$*$}}}
\def\Eval#1{\llbracket{#1}\rrbracket}
\def\Vec{\overrightarrow}

\def\Tilde{\widetilde}
\def\Break{\hfil\break\hbox{}}

\title{Equivalence of
Intuitionistic Inductive Definitions and 
Intuitionistic Cyclic Proofs under Arithmetic}
\author{
Stefano Berardi (Torino University)
\and  
Makoto Tatsuta (National Institute of Informatics)
}
\date{}

\maketitle

\begin{abstract}

A cyclic proof system gives us another way of
representing inductive definitions and efficient proof search.
In 2011 Brotherston and Simpson conjectured the equivalence
between the provability of the classical cyclic proof system and that
of the classical system of Martin-Lof's inductive definitions.
 This paper studies the conjecture for intuitionistic logic.
 This paper first points out that the countermodel of FOSSACS 2017 paper 
by the same authors
shows the conjecture for intuitionistic logic is false in general.
Then this paper shows the conjecture for intuitionistic logic is true under
arithmetic, namely, the provability of the intuitionistic
cyclic proof system is the same as that of the
intuitionistic system of Martin-Lof's inductive definitions when
both systems contain Heyting arithmetic HA.
 For this purpose, this paper also shows 
that HA proves Podelski-Rybalchenko theorem for induction and
Kleene-Brouwer theorem for induction.
These results immediately give another proof to 
the conjecture under arithmetic for classical logic shown in LICS 2017 paper by the same authors.

\end{abstract}

\section{Introduction}

An inductive definition is a way to define a predicate by an expression which may contain the predicate itself.
The predicate is interpreted by the least fixed point of the defining equation. 
Inductive definitions are important in computer science, since they can define useful recursive data structures such as lists and trees. 
Inductive definitions are important also in mathematical logic, since
they increase the proof theoretic strength.
Martin-L\"of's system of inductive definitions given in \cite{Martin-Lof-1971} is 
one of the most popular systems of inductive definitions.
This system has production rules for an inductive predicate,
and the production rule determines the introduction rules and the elimination rules for the predicate.

Brotherston and Simpson
\cite{Brotherston-phd,Brotherston11a} proposed an alternative formalization
of inductive definitions, called a cyclic proof system.
A proof, called {\em a cyclic proof}, is defined by proof search,
going upwardly in a proof figure.
If we encounter the same sequent (called a {\em bud})
as some sequent we already passed (called a {\em companion})
we can stop.
The induction rule is replaced by a case rule, for this purpose.
The soundness is guaranteed by 
some additional condition, called a {\em global trace condition},
which can show
the case rule decreases some measure of a bud from that of the companion.
In general, for proof search,
a cyclic proof system can find an induction formula in a more efficient way 
than Martin-L\"of's system,
since a cyclic proof system does not have to choose fixed induction formulas
in advance.
A cyclic proof system enables us efficient implementation of
theorem provers with inductive definitions \cite{Brotherston05,Brotherston08,Brotherston11,Brotherston12}.

Brotherston and Simpson \cite{Brotherston11a}
investigated
Martin-L\"of's system $\LKID$ of inductive definitions in classical logic
for the first-order language,
and
the cyclic proof system $\CLKIDomega$ for the same language,
showed the provability of
$\CLKIDomega$ includes that of $\LKID$,
and conjectured the equivalence.

By 2017, the equivalence was left an open question.
In general, the conjecture was proved to be false in \cite{Berardi17}, 
by showing a counterexample.
However, if we restrict both systems to 
only the natural number inductive predicate and 
add Peano arithmetic to both systems,
the conjecture was
proved to be true in \cite{Simpson17},
by internalizing a cyclic proof in $\rm ACA_0$ and using some results
in reverse mathematics.
\cite{Berardi17a} proved that
if we add Peano arithmetic to both systems,
$\CLKIDomega$ and $\LKID$ are equivalent,
namely the conjecture is true under arithmetic,
by showing arithmetical Ramsey theorem and 
Podelski-Rybalchenko theorem for induction.

This paper studies the conjecture for intuitionistic logic, namely,
the provability of the intuitionistic
cyclic proof system, called $\CLJIDomega$, is the same as that of the
intuitionistic system of Martin-Lof's inductive definitions, called $\LJID$.
This question is theoretically interesting, and
answers will potentially give new techniques of theorem proving by cyclic
proofs to type theories with inductive types and program
extraction by constructive proofs.

 This paper first points out that the countermodel of \cite{Berardi17}
also shows
the conjecture for intuitionistic logic is false in general.
 Then this paper shows the conjecture for intuitionistic logic is true under
arithmetic, namely, the provability of $\CLJIDomega$
is the same as that of $\LJID$,
when both systems contain Heyting arithmetic $\HA$.
Note that
a counterexample in \cite{Berardi17} does not work for a system
that contains $\HA$.

We explain main ideas of this paper.
The proof transformation given in \cite{Berardi17a} is intuitionistic, 
so we can use it for our purpose.
Only thing is to intuitionistically show $\Ind(>_\Pi)$.
\cite{Berardi17a} uses classical logic to show arithmetical Ramsey theorem
and Podelski-Rybalchenko theorem for induction,
and we cannot use this technique in $\HA$.
In order to show it we will take the following steps:

(1) For each $\pi \in B$, there is $n$ such that $\Ind(>_\pi^n)$.
($(\ )^n$ denotes the $n$-time composition.)

(2) Finiteness of path relations $\{ >_\pi \ |\ \pi \in B \}$.

(3) Kleene-Brouwer theorem for induction.

(4) Podelski-Rybalchenko theorem for induction.

The global trace condition gives (1). (4) is proved by (3).
Combining (2) and (4), we will obtain $\Ind(>_\Pi)$.
The places we need arithmetic are the proofs of (3) and (4), since
they use sequences of numbers.
 The claims (1) and (2) can be easily shown in almost the same way 
as \cite{Berardi17a}.
We will show the claim (3)
by refining an ordinary proof of Kleene-Brouwer theorem for orders.
We will show the claim (4)
by using Erd\"os trees and (3).

The results of this paper immediately give another proof 
to the conjecture under arithmetic 
for classical logic shown in \cite{Berardi17a}
by using the fact $\Gamma \prove_{\CLKIDPA} \Delta$ implies
$E,\Gamma,\neg\Delta \prove_{\CLJIDHA}$ for some finite set $E$
of excluded middles.

There are not papers that study the conjecture for intuitionistic logic or
Kleene-Brouwer theorem for induction in intuitionistic first-order logic.
For Podelski-Rybalchenko theorem for induction,
\cite{Berardi15} intuitionistically showed it but they used second-order logic.

Section 2 describes Brotherston-Simpson conjecture.
Section 3 defines $\CLJIDHA$ and $\LJIDHA$.
Section 4 explains main ideas.
Section 5 proves Kleene-Brouwer theorem for induction and 
Podelski-Rybalchenko theorem for induction.
Section 6 discusses proof transformation and the main theorem.
We conclude in Section 7.

\section{Brotherston-Simpson Conjecture for Intuitionistic Logic}

\subsection{Intuitionistic Martin-L\"of's Inductive Definition System $\LJID$}

An intuitionistic Martin-L\"of's inductive definition system, called $\LJID$,
is defined as the system obtained from
classical Martin-L\"of's inductive definition system $\LKID$
defined in \cite{Brotherston11a}
by restricting every sequent to intuitionistic sequents and replacing
$(\imp L)$, $(\lor R)$, and $(\Ind\ P_j)$ by
{\prooflineskip
\[
\infer[(\imp L)]{\Gamma, F \imp G \prove \Delta}{
	\Gamma \prove F
	&
	\Gamma,G \prove \Delta
}
\qquad
\infer[(\lor R_l)]{\Gamma \prove F \lor G}{
	\Gamma \prove F
}
\qquad
\infer[(\lor R_r)]{\Gamma \prove F \lor G}{
	\Gamma \prove G
}
\\
\infer[(\Ind\ P_j)]{\Gamma,P_j\Vec u \prove F_j\Vec u}{
\hbox{minor premises}
}
\]
}%
where 
the minor premises are the same as the minor premises 
of $(\Ind\ P_j)$ in $\LKID$ (page 9 of \cite{Brotherston11a}).
Note that we replace these rules because
their formalization in $\LKID$
does not work
for intuitionistic logic.

\subsection{Cyclic Proof System $\CLJIDomega$}

An intuitionistic cyclic proof system, called $\CLJIDomega$,
is defined as the system obtained from
classical cyclic proof system $\CLKIDomega$
defined in \cite{Brotherston11a}
by restricting every sequent to intuitionistic sequents and
replacing $(\imp L)$ and $(\lor R)$ in the same way as $\LJID$.
Note that the {\em global trace condition} in $\CLJIDomega$ 
is the same as that in $\CLKIDomega$ (Definition 5.5 of \cite{Brotherston11a}).

\subsection{Brotherston-Simpson Conjecture}

Brotherston-Simpson conjecture (the conjecture 7.7 in \cite{Brotherston11a}) is
that the provability of $\LKID$ is the same as that of $\CLKIDomega$.
In general, the conjecture was proved to be false in \cite{Berardi17}, 
by showing a counterexample.
\cite{Berardi17a} proved
that the conjecture is true
for any inductive predicates
with their stage-number inductive predicates,
if we add arithmetic to both systems.

This paper studies an intuitionistic version of the conjecture,
namely the equivalence between $\CLJIDomega$ and $\LJID$.

The counterexample given in \cite{Berardi17} also shows that
the equivalence between $\CLJIDomega$ and $\LJID$ does not hold in general,
because the proof of the statement $H$ in \cite{Berardi17} is actually in $\CLJIDomega$,
and $\LJID$ does not prove $H$ since $\LKID$ does not prove $H$.
This gives us the following theorem.
\begin{Th}
There are some signature and some set of production rules such that
the provability of $\CLJIDomega$ is not the same as
that of $\LJID$.
\end{Th}

This paper will show that
the same results as \cite{Berardi17a} for intuitionistic logic,
namely,
the provability of $\LJID$ is the same as that of $\CLJIDomega$
if we add Heyting arithmetic to both systems.
This means
that the conjecture is true for intuitionistic logic
under arithmetic.

\section{Addition of Heyting Arithmetic}

In this section,
we define systems $\CLJIDHA$ and $\LJIDHA$.

\begin{Def}\rm
$\CLJIDHA$ and $\LJIDHA$
are defined to be obtained from $\CLJIDomega$ and $\LJID$
by adding Heyting arithmetic.
Namely, we add constants and function symbols $0, s, +, \times$,
the inductive predicate symbol $\N$, the productions for $\N$,
and Heyting axioms:
\[
\infer{\N0}{}
\qquad
\infer{\N s x}{\N x}
\qquad
\prove \N x \imp s x \ne 0, \qquad
\prove \N x \land \N y \imp s x=s y \imp x=y, \\
\prove \N x \imp x + 0 = x, \qquad
\prove \N x \land \N y \imp x + s y = s(x+y), \\
\prove \N x \imp x \times 0 = 0, \qquad
\prove \N x \land \N y \imp x \times s y = x \times y + x.
\]
\end{Def}

We define $x < y$ by $\exists z.x+s z=y$
and $x \le y$ by $x=y \lor x<y$.

We can assume some coding of a sequence of numbers by a number in Heyting arithmetic, because
the definitions on pages 115--117 of \cite{Shoenfield67} work also in $\HA$.
We write $\< t_0,\ldots,t_n \>$
for
the sequence of $t_0,\ldots,t_n$.
In particular, we write $\<\ \>$ for the empty sequence.
We define the $i$-th element of $t_0,\ldots,t_n$ as $t_i$.
We also write $|t|$, and $(t)_u$ for 
the length of the sequence $t$, and
the $u$-th element of the sequence $t$ respectively.
Note that $|\< t_0,\ldots,t_n \>|=n+1$ and
$(\< t_0,\ldots,t_n \>)_i=t_i$.
We write $*$ for the concatenation operation of sequences.

We call an atomic formula an {\em inductive atomic formula}
when its predicate symbol is an inductive predicate symbol.
For a predicate $P$, we sometimes write $t \in P$ for $P(t)$.

\section{Main Ideas}

In \cite{Berardi17a},
a given proof in $\CLKIDPA$ is transformed into 
a proof in $\LKIDPA$.
The main construction of the proof in $\LKIDPA$ is summarized by
\[
\infer{J}{
\infer[(2)]{Gx_0x}{
	\Ind(>_\Pi)
	&
	\infer{Gx_0x}{
	\infer*[\hbox{each subproof}]{J_2}{
	\infer[(1)]{\Ineq \imp J_1}{
	\infer{J_1}{
	\infer{Gy_0y}{
		\infer{\<x_0,x\> >_\Pi \<y_0,y\>}{(1) \Ineq}
		&
		(2) Hx_0x
	}}}}}
}}
\]
One of the key points in this proof construction is
the proof of $\Ind(>_\Pi)$.

The same construction works for $\CLJIDHA$ and $\LJIDHA$ except
$\Ind(>_\Pi)$, since the proof uses only intuitionistic sequents
when the goal sequent is an intuitionistic sequent.

In order to intuitionistically show $\Ind(>_\Pi)$, we will take the following steps:

(1) For each $\pi \in B$, there is $n$ such that $\Ind(>_\pi^n)$.

(2) Finiteness of path relations $\{ >_\pi \ |\ \pi \in B \}$.

(3) Kleene-Brouwer theorem for induction.

(4) Podelski-Rybalchenko theorem for induction.

The global trace condition gives (1). (4) is proved by (3).
Combining (2) and (4), we will obtain $\Ind(>_\Pi)$.
The places we need arithmetic are the proofs of (3) and (4), since
they use sequences of numbers.

The claim (1) can be shown in the same way as \cite{Berardi17a}
since that paper did not use classical logic for proving (1).
The claim (2) can be easily shown also in intuitionistic logic 
by using iteration to the least point.

For (3),
we will show
Kleene-Brouwer theorem for induction,
which states that 
if we have 
both 
induction principle for a lifted tree (namely $\<u\>*T$ for some tree $T$) 
with respect to the one-step extension relation
and induction principle for relations on children, then
we have induction principle for the Kleene-Brouwer relation.
We can prove it by refining an ordinary proof of Kleene-Brouwer theorem
for orders.

For (4),
we will show Podelski-Rybalchenko theorem for induction
stating that
if transition invariant $>_\Pi$
is a finite union of relations $>_\pi$ such that 
each $\Ind(>_\pi^n)$ is provable for some $n$,
and each $(>_\pi)$ is decidable,
then
$\Ind(>_\Pi)$ is provable.

First each $\Ind(>_\pi)$ is obtained by $\Ind(>_\pi^n)$.
Next by the decidability of each $(>_\pi)$, we can assume
all of $(>_\pi)$ are disjoint to each other.
For simplicity, we explain the idea of our proof 
for well-foundedness instead of induction principle.

Assume the relation $>_\Pi$ has some decreasing transitive sequence
\[
u_1 >_\Pi u_2 >_\Pi u_3 >_\Pi \ldots
\]
in order to show contradiction,
where
a sequence is called {\em transitive} when $u_i >_{\Pi} u_j$ for any $i<j$.

We say the edge from $u$ to $v$ is of color $R_i$ when $u >_{R_i} v$.
A sequence is called {\em monotonically-colored} when
for any element there is a color such that
the edge from the element to any element after it in the sequence 
has the same color.
Define $\MS$ as the set of monotonically-colored finite sequences.
It will be shown to be well-founded with the one-step extension relation.
A set of sequences beginning with the same element in some tree is called
a {\em lifted tree}.
For a decreasing transitive sequence $x$ of $U$,
a lifted tree $T \in U^{<\omega}$ is called an {\em Erd\"os tree} of $x$ when
the elements of $x$ are the same as elements of elements of $T$,
every element of $T$ is monotonically-colored,
and the edges from a parent to its children have different colors.
Let $\ET$ be a function that returns an Erd\"os tree.
Then we consider
\[
\ET(\<u_1\>), \ET(\<u_1,u_2\>), \ET(\<u_1,u_2,u_3\>), \ldots.
\]

Define $\MS_{\<r\>}$ as the set of sequences beginning with $r$ in $\MS$.
Define $>_{\KB1,r}$ as the Kleene-Brouwer relation for 
the lifted tree $\MS_{\<r\>}$
and some left-to-right-decreasing relation on children of the lifted tree.
Define $>_{\KB2,r}$ as the Kleene-Brouwer relation for 
the lifted tree $\MS_{\<r\>}$
and some right-to-left-decreasing relation on children of the lifted tree.
By Kleene-Brouwer theorem, $(>_{\KB1,r})$ and $(>_{\KB2,r})$ are well-founded.
Define $\ET2(\<u_1,\ldots,u_n\>)$ as 
the $(>_{\KB2,u_1})$-sorted sequence of elements in $\ET(\<u_1,\ldots,u_n\>)$.
Then consider
\[
\ET2(\<u_1\>), \ET2(\<u_1,u_2\>), \ET2(\<u_1,u_2,u_3\>), \ldots.
\]

Define $>_{\KB,r}$ as the Kleene-Brouwer relation for 
$>_{\KB1,r}$ and 
the set of $(>_{\KB2,r})$-sorted finite sequences of elements in $\MS_{\<r\>}$.
This definition is a key idea.
By this definition, we have
\[
\ET2(\<u_1\>) >_{\KB,u_1} \ET2(\<u_1,u_2\>) >_{\KB,u_1} \ET2(\<u_1,u_2,u_3\>) >_{\KB,u_1} \ldots.
\]
Since $(>_{\KB,u_1})$ is well-founded by Kleene-Brouwer theorem, 
we have contradiction.

In general we need classical logic to derive induction principle from well-foundedness,
but the idea we have explained will work well for showing induction principle in intuitionistic logic.

\section{$\HA$-Provable Podelski-Rybalchenko Theorem for Induction}

This section will prove
Podelski-Rybalchenko theorem for induction, inside Heyting arithmetic $\HA$.
First we will prove Kleene-Brouwer theorem for induction, inside $\HA$.
Next we will show induction for
the set $\MS$ of monotonically-colored subsequences.
Then by applying Kleene-Brouwer theorem to a part of $\MS$ and some orders
$>_{u,\Left}$ and $>_{u,\Right}$,
we will obtain two Kleene-Brouwer relations $>_{\KB1,r}$
and $>_{\KB2,r}$ and show their induction principle.
Then by applying Kleene-Brouwer theorem to 
some lifted tree determined by $>_{\KB2,r}$ and
the relation $>_{\KB1,r}$, 
we will obtain a Kleene-Brouwer relation $>_{\KB,r}$ and show its induction principle.
Then we will show
induction for decreasing transitive sequences
is reduced to
induction for Erd\"os trees with the relation $>_{\KB,r}$.
Since Erd\"os trees are in the lifted tree,
by combining them, we will prove
Podelski-Rybalchenko theorem for induction.

We write $>_R$ or $>$ for a binary relation.
We write $<_R$ for the binary relation of the inverse of $>_R$.
We write $y <_R x \in X$ for $y <_R x \land y \in X$.
We write $x \in \sigma$ when $x$ is an element of the sequence $\sigma$.
We write $U^{<\omega}$ for the set of finite sequences of elements in $U$.
For a set $S$ of sequences,
we write $\<u\>*S$ for $\{ \<u\>*\sigma \ |\ \sigma \in S\}$.
For a set $U$ and a binary relation $>_R$ for $U$, 
the {\em induction principle} for $(U,>_R)$ is defined as
\[
\Ind(U,>_R,F) \equiv \forall x \in U((\forall y<_R x \in U.Fy) \imp Fx)
	\imp \forall x \in U.Fx, \\
\Ind(U,>_R) \equiv \Ind(U,>_R,F) \ \hbox{(for every formula $Fx$)}.
\]

For a set $U$
a set $T$ is called a {\em tree} of $U$ if
$T \subseteq U^{<\omega}$ and $T$ is nonempty and closed under prefix operations.
Note that the empty sequence is a prefix of any sequence.
As a graph,
the set of nodes is $T$ and
the set of edges is $\{ (x,y) \in T^2 \ |\ y = x *\<u\> \}$.
We call a set $T \subseteq U^{<\omega}$ a {\em lifted tree} of $U$ when
there are a tree $T' \subseteq U^{\omega}$ and $r \in U$ such that
$T=\<r\>*T'$.
We define $\LiftedTree(T,U)$ as
a first-order formula that means $T$ is a lifted tree of $U$.

For $x,y \in U^{<\omega}$ we define the one-step extension relation $x >_\Ext y$ if
$y = x * \< u \>$ for some $u$.
For a set $T \subseteq U^{<\omega}$ and $\sigma \in U^{<\omega}$,
we define $T_\sigma$ as
$\{ \rho \in T \ |\ \rho = \sigma * \rho' \}$.
Note that $T_\sigma$ is a subset of $T$.
For a nonempty sequence $\sigma$,
we define $\First(\sigma)$ and $\Last(\sigma)$ 
as the first and the last element of $\sigma$ respectively.

The next lemma shows induction implies 
$x\not>x$.
\begin{Lemma}\label{lemma:ind-ne}
If $\HA \prove \Ind(U,>)$, then $\HA \prove \forall x,y \in U(y < x \imp y \ne x)$.
\end{Lemma}

{\em Proof.}
Fix $x,y \in U$ and 
assume $y<x$ and $y=x$ in order to show contradiction.

Define
$Fz$ be $z \ne x$.
Then we can show
\[
\HA \prove \forall z\in U((\forall w<z\in U.Fw) \imp Fz)
\]
by case analysis for $z \ne x \lor z = x$ as follows.
In the first case $z \ne x$, $Fz$.
In the second case, by taking $w$ to be $x$ in $\forall w<z\in U.Fw$, 
we have $Fx$.

By $\Ind(U,>)$, $\forall z\in U.Fz$. By taking $z$ to be $x$, we have contradiction.
$\Box$

\begin{Def}[Kleene-Brouwer Relation]\rm
For a set $U$,
a lifted tree $T$ of $U$, 
and 
the set of binary relations $>_u$ on $U$ for every $u \in U$,
we define the {\em Kleene-Brouwer relation} $>_\KB$ for $T$ and 
$\{ (>_u) \ |\ u \in U\}$ as follows:
for $x,y \in T$, $x >_\KB y$ if
(1) $x = z * \<u, u_1 \> * w_1$,
$y = z * \<u, u_2 \> * w_2$, and $u_1 >_u u_2$
for some $z,u,u_1,w_1,u_2,w_2$,
or (2) $y = x * z$ for some $z \ne \<\ \>$.

When $(>_u)$ is some fixed $(>)$ for all $u$, 
for simplicity we call the relation $(>_\KB)$
the Kleene-Brouwer relation for $T$ and $>$.
\end{Def}
Note that $(>_\KB)$ is a relation on $T$.
This Kleene-Brouwer relation is slightly different from
ordinary Kleene-Brouwer order for the following points:
it creates a relation instead of an order,
it uses a set of relations indexed by an element,
and
it is defined for a lifted tree instead of a tree
(in order to use indexed relations).

The next theorem shows
Kleene-Brouwer theorem for induction, which states that
if we have 
both 
induction principle for a lifted tree with respect to the extension relation
and induction principle for relations on children, then
we have induction principle for the Kleene-Brouwer relation.
\begin{Th}[Kleene-Brouwer Theorem for Induction]\label{th:KB}
If 
$\HA \prove \LiftedTree(T,U)$,
$\HA \prove \Ind(T,>_\Ext)$ and $\HA \prove \forall u\in U.\Ind(U,>_u)$,
then $\HA \prove \Ind(T,>_\KB)$.
\end{Th}

{\em Proof.}
By induction on $(T,>_\Ext)$, we will show 
$
\forall \sigma \in T.\Ind(T_\sigma,>_\KB).
$
After we prove it, we can take $\sigma$ to be $\<\ \>$ to show the theorem,
since $T_{\<\ \>}=T$.

Fix $\sigma \in T$ in order to show $\Ind(T_\sigma,>_\KB)$.
Note that we can use induction hypothesis for every $\sigma*\<u\> \in T$:
\begin{eq}0
\Ind(T_{\sigma*\<u\>},>_\KB).
\end{eq}%

Assume 
\begin{eq}1
\forall x \in T_\sigma((\forall y <_\KB x \in T_\sigma. Fy) \imp Fx)
\end{eq}%
in order to show $\forall x \in T_\sigma.Fx$.
For simplicity we write $F(X)$ for $\forall x\in X.Fx$. 
Let $Gu \equiv F(T_{\sigma * \< u \>})$.
By $\Ind(U,>_{\Last(\sigma)})$ we will show the following claim.

Claim: $\forall u \in U.Gu$.

Fix $u \in U$ in order to show $Gu$.

By IH for $v$ with $>_{\Last(\sigma)}$
we have
\begin{eq}2
v <_{\Last(\sigma)} u \imp F(T_{\sigma * \< v \>}).
\end{eq}%

We can show
\begin{eq}3
\forall x\in T_{\sigma*\<u\>}((\forall y <_\KB x \in T_{\sigma*\<u\>}.Fy)
\imp
(\forall y <_\KB x \in T_\sigma. Fy))
\end{eq}%
as follows.
Fix $x\in T_{\sigma*\<u\>}$,
assume
\begin{eq}4
\forall y <_\KB x \in T_{\sigma*\<u\>}.Fy
\end{eq}%
and
assume $y <_\KB x \in T_\sigma$ in order to show $Fy$.
By definition of $>_\KB$,
we have
$y \in T_{\sigma * \< v \>}$ for some $v <_{\Last(\sigma)} u$, or $y \in T_{\sigma * \< u \>}$.
In the first case, $Fy$ by \eqref2.
In the second case, $Fy$ by \eqref4.
Hence we have shown \eqref3.

Combining \eqref3 with \eqref1, we have
\begin{eq}5
\forall x \in T_{\sigma*\<u\>}((\forall y <_\KB x \in T_{\sigma * \<u\>}.Fy) \imp F(x)).
\end{eq}%

By IH \eqref0 for $\sigma * \< u \>$, we have $\Ind(T_{\sigma * \< u \>}, >_\KB)$,
namely,
\begin{eq}6
\forall x \in T_{\sigma*\<u\>}((\forall y <_\KB x \in T_{\sigma * \<u\>}.Fy) \imp Fx) \imp \forall x \in T_{\sigma*\<u\>}.Fx.
\end{eq}%
By \eqref6\eqref5, 
$F(T_{\sigma * \<u\>})$. 
Hence we have shown the claim.

If $y <_\KB \sigma \in T_\sigma$, 
we have $y \in T_{\sigma * \<u\>}$ for some $u$,
since $y <_\KB \sigma$ implies $y \ne \sigma$ by definition of $\KB$
and Lemma \ref{lemma:ind-ne} for $>_u$.
By the claim, $Fy$. Hence 
\begin{eq}7
\forall y<_\KB \sigma \in T_\sigma.Fy.
\end{eq}%

By letting $x:=\sigma$ in \eqref1, we have
$(\forall y<_\KB \sigma \in T_\sigma.Fy) \imp F\sigma$.
By \eqref7, $F\sigma$.
Combining it with the claim, $\forall x\in T_\sigma.Fx$.
$\Box$

\begin{Def}\rm
For a set $U$ and a relation $>$ for $U$,
we define the set $\DS(U,>)$ 
of decreasing sequences 
as $\{ \<x_0, \ldots, x_{n-1}\> \ |\ n \ge 0,
	x_i \in U, \forall i<n-1.(x_i > x_{i+1}) \}$.

We define the set $\DT(U,>)$ 
of decreasing transitive sequences 
by $\{ \<x_0, \ldots, x_{n-1}\> \ |\ n\ge 0,
	x_i \in U, \forall i(\forall j \le n-1.(i < j \imp x_i > x_j)) \}$.

We define $>_{R_1\cup\ldots\cup R_k}$ as the union of $>_{R_i}$ for all $1 \le i \le k$.
We define $>_{R_1+\ldots+R_k}$ as the disjoint union of $>_{R_i}$ for all $1 \le i \le k$.
(Whenever we use it, we implicitly assume 
the disjointness is provable in $\HA$.)

We define $\Monoseq_{R_1,\ldots,R_k}(x)$ to hold when
$x = \<x_0,\ldots,x_{n-1}\> \in \DT(U,>_{R_1+  \ldots+  R_k})$ and
$\forall i<n-1.(\forall j \le n-1.(i < j \imp \Land_{1 \le l \le k}(
x_i >_{R_l} x_{i+1} \imp x_i >_{R_l} x_j)))$.
Note that $n$ may be 0.

We define $\MS$
as
$\{ x \in \DT(U,>_{R_1+  \ldots+  R_k}) \ |\ \Monoseq_{R_1,\ldots,R_k}(x) \}$.
\end{Def}
$\MS$ is the set of monotonically-colored finite sequences.
Note that
$\MS_{\<r\>}$ is 
a subset of $\MS$ and 
a lifted tree of $U$ for any $r \in U$.

\begin{Def}\rm
For a relation $>_{R_i}$ on $U_i$ for $1 \le i \le k$,
we define a relation $>_{R_1 \times \ldots \times R_k}$ on 
$U_1 \times \ldots \times U_k$ by:
$(x_1,\ldots,x_k) >_{R_1 \times \ldots \times R_k} (y_1,\ldots,y_k)$ if
there is some $i$ such that 
$x_i >_{R_i} y_i$ and $x_j = y_j$ for all $j \ne i$.
\end{Def}

The next lemma shows induction for cartesian product.
\begin{Lemma}\label{lemma:ind-product}
If $\HA \prove \Ind(U_i,>_{R_i})$ for $1 \le i \le k$, 
then $\HA \prove 
\Ind(U_1 \times \ldots \times U_k,>_{R_1 \times \ldots \times R_k})$.
\end{Lemma}

{\em Proof.}
First we will show the case $k=2$.

For simplicity, we write $U$ for $U_1 \times U_2$ and
$>_\times$ for $>_{R_1 \times  R_2}$.

Assume
\begin{eq}1
\forall x \in U((\forall y<_\times x \in U.Fy) \imp Fx)
\end{eq}%
in order to show $\forall x\in U.Fx$.

Define $Gx_1 \equiv \forall x_2\in U_2.F(x_1,x_2)$.
We will show
\begin{eq}2
\forall x_1\in U_1((\forall y_1<_{R_1}x_1\in U_1.Gy_1) \imp Gx_1).
\end{eq}%
Fix $x_1\in U_1$ and assume
\begin{eq}3
\forall y_1<_{R_1}x_1\in U_1.Gy_1
\end{eq}%
in order to show $Gx_1$.
We will show
\begin{eq}4
\forall x_2\in U_2((\forall y_2<_{R_2}x_2.F(x_1,y_2)) \imp F(x_1,x_2)).
\end{eq}%
Fix $x_2\in U_2$ and assume
\begin{eq}5
\forall y_2<_{R_2}x_2.F(x_1,y_2)
\end{eq}%
in order to show $F(x_1,x_2)$.
We will show
\begin{eq}6
\forall y<_\times (x_1,x_2)\in U.Fy.
\end{eq}%
Fix $y<_\times (x_1,x_2)\in U$ in order to show $Fy$.
Let $(y_1,y_2)$ be $y$.
Consider cases by $y<_\times (x_1,x_2)$.

Case 1. $y_1 = x_1$ and $y_2<_{R_2}x_2$.

By taking $y_2$ to be $y_2$ in \eqref5, $F(x_1,y_2)$. Hence $Fy$.

Case 2. $y_1<_{R_1}x_1$ and $y_2=x_2$.

By taking $y_1$ to be $y_1$ in \eqref3, $Gy_1$. 
Hence $\forall x_2\in U_2.F(y_1,x_2)$.
By taking $x_2$ to be $x_2$ in it, $F(y_1,x_2)$. Hence $Fy$.

In both cases, $Fy$. Hence we have shown \eqref6.
By taking $x$ to be $(x_1,x_2)$ in \eqref1, we have $F(x_1,x_2)$.
Hence we have shown \eqref4.
By $\Ind(U_2,>_{R_2})$ for $\lambda x_2.F(x_1,x_2)$,
we have $\forall x_2\in U_2.F(x_1,x_2)$.
Hence $Gx_1$.
Hence we have shown \eqref2.
By $\Ind(U_1,>_{R_1})$ for $G$, we have
$\forall x_1\in U_1.Gx_1$.
Hence $\forall x\in U.Fx$.

We have shown the case $k=2$.

Next we will show the case $k>2$.
We use induction on $k$ to show the claim.
By IH, we have 
$\Ind(U_1 \times \ldots \times U_{k-1},>_{R_1 \times \ldots \times R_{k-1}})$.
By using the case $k=2$ for it and $\Ind(U_k,>_{R_k})$,
we have
$\Ind(U_1 \times \ldots \times U_{k},>_{(R_1 \times \ldots \times R_{k-1}) \times R_k})$.
Since $(>_{(R_1 \times \ldots \times R_{k-1}) \times R_k})$ is
$(>_{R_1 \times \ldots \times R_k})$,
we have the claim.
$\Box$

The next lemma shows that
induction principle for each relation
implies
induction principle for monotonically-colored sequences.
This lemma can be proved by refining
Lemma 6.4 (1) of \cite{Berardi15} 
from second-order logic to first-order logic.

\begin{Lemma}\label{lemma:MS}
If $\HA \prove \Ind(\DT(U,>_{R_{i}}), >_\Ext)$ for all $ 1 \le i \le k$,
then $\HA \prove \forall r\in U.\Ind(\MS_{\<r\>},>_\Ext)$.
\end{Lemma}

{\em Proof.}
Fix $r \in U$ in order to show $\Ind(\MS_{\<r\>},>_\Ext)$.

Assume
\begin{eq}1
\forall \sigma \in \MS_{\<r\>}((\forall \rho <_\Ext \sigma \in \MS_{\<r\>}.F\rho) \imp F\sigma)
\end{eq}%
in order to show $\forall \sigma \in \MS_{\<r\>}.F\sigma$.

For $1 \le i \le k$,
define
\[
\Seq_i(\sigma) = \<x_{n_1}, \ldots, x_{n_m}\>
\]
where 
$\sigma = \<x_1,\ldots,x_n\>$,
$\{x_{n_1}, \ldots, x_{n_m}\} = \{x_j \in \sigma \ |\ x_j >_{R_i} x_{j+1} \}$,
and
$n_1<_N \ldots<_N n_m$ for the natural number order $>_N$.
Formally $\Seq_i(\sigma)=\rho$ is an abbreviation for
some $\HA$-formula $F(\sigma,\rho)$.
Note that $\Seq_i(\sigma)$ may be $\<\ \>$.

For simplicity we write $\DT_k$ for
 $\DT(U,>_{R_1}) \times \ldots \times \DT(U,>_{R_k})$.

We define $\Ext^k$ for $\DT_k$ by
\[
(x_1,\ldots,x_k) >_{\Ext^k} (y_1,\ldots,y_k)
\]
where for some $1 \le i \le k$, $x_i >_\Ext y_i$ and $x_j=y_j$ for all $j \ne i$.
Note that the set of elements of $\sigma$ is the union of
the sets of $\Seq_{i}(\sigma)$ $(1 \le i \le k)$ and $\{ \Last(\sigma) \}$.

Define
\[
G(x_1,\ldots,x_k) \equiv \forall\sigma\in \MS_{\<r\>}(
	(1 \le \forall i \le k.\Seq_{i}(\sigma)=x_{i}) \imp F\sigma).
\]

We write $\Vec x$ for $(x_1,\ldots,x_k)$.
We will show
\begin{eq}2
\forall \Vec x \in \DT_k((\forall \Vec y <_{\Ext^k} \Vec x \in \DT_k.G(\Vec y)) \imp G(\Vec x)).
\end{eq}%
Fix $\Vec x$ and
assume
\begin{eq}3
\forall \Vec y <_{\Ext^k} \Vec x \in \DT_k.G(\Vec y)
\end{eq}%
in order to show $G\Vec x$.
Fix $\sigma \in \MS_{\<r\>}$ and assume $\Seq_{i}(\sigma) = x_{i}$ for all $1 \le i \le k$
in order to show $F\sigma$.

We can show $\forall\rho <_\Ext \sigma \in \MS_{\<r\>}.F\sigma$ as follows.
Assume $\rho <_\Ext \sigma$.
Let $\rho = \sigma * \<u\>$.
Then $\Last(\sigma) >_{R_{i}} u$ for some $1 \le i \le k$.
Then
\[
\Seq_i(\rho) = \Seq_i(\sigma) *\<\Last(\sigma)\>, \qquad
\Seq_j(\rho) = \Seq_j(\sigma) \ \ (\forall j \ne i).
\]
Hence $(\Seq_1(\rho),\ldots,\Seq_k(\rho)) <_{\Ext^k}
(\Seq_1(\sigma),\ldots,\Seq_k(\sigma))$, namely,
$(\Seq_1(\rho),\ldots,\Seq_k(\rho)) <_{\Ext^k} \Vec x$.
By \eqref3, $G(\Seq_1(\rho),\ldots,\Seq_k(\rho))$. Hence $F\rho$.
Hence we have shown $\forall\rho <_\Ext \sigma \in \MS_{\<r\>}.F\sigma$.

By \eqref1, $F\sigma$.
Hence we have shown $G\Vec x$.
Hence we have shown \eqref2.

By Lemma \ref{lemma:ind-product} for 
$\Ind(\DT(U,>_{R_{i}}), >_\Ext)$ for all $ 1 \le i \le k$,
we have $\Ind(\DT_k, >_{\Ext^k})$.
By it and \eqref2,
we have
\[
\forall\Vec x \in \DT_k.G\Vec x.
\]
For every $\sigma \in \MS_{\<r\>}$,
by letting $x_i=\Seq_i(\sigma)$ for all $1 \le i \le k$,
we have $G\Vec x$, and hence we have $F\sigma$.
$\Box$

Next we create
Kleene-Brouwer relations $>_{\KB1,r}$ and $>_{\KB2,r}$ for
monotonically-colored sequences beginning with $r$.
Then we consider
the set of
$(>_{\KB2,r})$-sorted
finite sequences of monotonically-colored finite sequences beginning with $r$.
It is a lifted tree.
Then, by induction principle for $\MS$,
the lifted tree is well-founded with the one-step extension relation.
The Kleene-Brouwer relation for the lifted tree and $>_{\KB1,r}$
gives us $>_{\KB,r}$ for the lifted tree.
Since an Erd\"os tree is in the lifted tree,
this will later show induction principle for Erd\"os trees.
\begin{Def}\rm
For $u \in U$,
we define $>_{u,\Left}$ for $U$ by:  $u_1 >_{u,\Left} u_2$ if
$u >_{R_j} u_1$,
$u >_{R_l} u_2$,
and $j<l$
for some $j,l$.

We define $>_{\KB1,r}$ for $\MS_{\<r\>}$
as the KB relation for $\MS_{\<r\>} \subseteq U^{<\omega}$ and $(>_{u,\Left}) \subseteq U^2$ for all $u \in U$.

For $u \in U$,
we define $>_{u,\Right}$ for $U$ by:  $u_1 >_{u,\Right} u_2$ if
$u_1 <_{u,\Left} u_2$.

We define $>_{\KB2,r}$ for $\MS_{\<r\>}$
as the KB relation for $\MS_{\<r\>} \subseteq U^{<\omega}$ and 
$(>_{u,\Right}) \subseteq U^2$ for all $u \in U$.

We define $>_{\KB,r}$ for $\DS(\MS_{\<r\>},>_{\KB2,r})_{\<\<r\>\>}$
as the KB relation for
$\DS(\MS_{\<r\>},>_{\KB2,r})_{\<\<r\>\>} \subseteq \MS_{\<r\>}^{<\omega}$ 
and $>_{\KB1,r}$.
\end{Def}
$>_{u,\Left}$ is 
the left-to-right-decreasing order of children of $u$ in 
some ordered tree of $U$ in which
the edge label $R_i$ is put to an edge $(x,y)$ such that $x >_{R_i} y$,
each parent has at most one child of the same edge label,
and children are ordered by their edge labels with $R_1 < \ldots < R_k$.
Similarly $>_{u,\Right}$ is 
the right-to-left-decreasing order of children of $u$ in the ordered tree.

\begin{Def}\rm
For $u \in U \subseteq N$, finite $T \subseteq \MS$
such that $\forall \rho \in T.\forall v \in \rho.(v>_{R_1+\ldots+R_k} u)$,
and for $\sigma \in T$,
we define the function $\Insert$ by:
\[
\Insert(u,T,\sigma) = \\ \qquad
\Insert(u,T,\sigma*\<v\>) {\rm\ if\ } 
\Last(\sigma) >_{R_i} u,
v = \mu v.(\sigma*\<v\> \in T \land \Last(\sigma) >_{R_i} v), \\
\qquad T \cup \{ \sigma*\<u\> \} {\rm\ otherwise},
\]
where $\mu v.F(v)$ denotes the least element $v$ with the natural number order
such that $F(v)$.
Formally $\Insert(u,T,\sigma)=T'$ is an abbreviation of
some $\HA$-formula $G(u,T,\sigma,T')$.
It is the same for $\ET$ below.

For $x \in \DT(U,>_{R_1+ \ldots+ R_k}) - \{\<\ \>\}$,
we define $\ET(x) \subseteq \MS$ by
\[
\ET(\<u\>) = \{ \<u\> \}, \\
\ET(x * \<u\>) = \Insert(u,\ET(x),\<\First(x)\>) {\ \rm if\ } x \ne \<\ \>.
\]
\end{Def}
Note that
$\Insert(u,T,\sigma)$
adds a new element $u$ to the set $T$ 
at some position below $\sigma$
to obtain a new set.
$\ET(x)$ is an Erd\"os tree obtained from the decreasing transitive sequence $x$.

The next lemma 
(1) states a new element is inserted at a leaf.
It is proved by induction on the number of elements in $T$.
The claim (2) states that edges from a parent to its children have different colors.
It is proved by induction on the length of $x$.
\begin{Lemma}\label{lemma:insert}
(1) For $u \in U$, 
$T \subseteq \MS$,
and
$\sigma \in T$,
if 
$u \notin \rho$ for all $\rho \in T$,
$\sigma = \<x_0,\ldots,x_{n-1}\>$, 
$x_i >_{R_j} x_{i+1}$ implies $x_i >_{R_j} u$ for all $i<n-1$,
and
$\Insert(u,T,\sigma)=T'$, then
there is some $\rho \in T_\sigma$ such that
$\rho*\<u\> \in \MS$,
$T'=T + \{\rho*\<u\>\}$,
and
$\rho*\<u\>$ is a maximal sequence in $T'$.

(2) If $\sigma*\<u,u_1\>*\rho_1, \sigma*\<u,u_2\>*\rho_2 \in \ET(x)$,
$u >_{R_i} u_1$, and $u >_{R_i} u_2$, then $u_1=u_2$.
\end{Lemma}

\begin{Def}\rm
For $x \in \DT(U,>_{R_1+ \ldots+ R_k}) - \{\<\ \>\}$, 
we define
\[
\ET2(x) \equiv \<x_0,\ldots,x_{n-1}\>
\]
where $\{x_0,\ldots,x_{n-1}\} = \ET(x)$ and 
$\forall i<n-1.(x_i >_{\KB2,\First(x)} x_{i+1})$.
\end{Def}
Note that $>_{\KB2,\First(x)}$ is a total order on $\ET(x)$ by Lemma \ref{lemma:insert} (2).
$\ET2(x)$ is the decreasing sequence of all nodes in the Erd\"os tree $\ET(x)$
ordered by $>_{\KB2,\First(x)}$.

The next lemma shows $\ET2$ is monotone.
It is the key property of reduction in Lemma \ref{lemma:ET-ind}.

\begin{Lemma}\label{lemma:ET}
$\HA \prove 
\forall r \in U.\forall x,y \in \DT(U,>_{R_1+ \ldots+ R_k})_{\<r\>}.
(x >_\Ext y \imp \ET2(x) >_{\KB,r} \ET2(y))$.
\end{Lemma}

{\em Proof.}
Fix $r \in U$ and $x,y \in \DT(U,>_{R_1+ \ldots+ R_k})_{\<r\>}$ and
assume $x >_\Ext y$.
Let $y = x * \<u\>$.
Then $\ET(y)=\Insert(u,\ET(x),\<r\>)$.
By Lemma \ref{lemma:insert} (1), we have $\sigma$ such that
$\ET(y) = \ET(x) + \{ \sigma*\<u\> \}$.
Then we have two cases:

Case 1. $\Last(\ET2(x)) >_{\KB2,r} \sigma*\<u\>$.

Then
$\ET2(y) = \ET2(x) *\<\sigma*\<u\>\>$.
By definition, $\ET2(x) >_{\KB,r} \ET2(y)$.

Case 2. 
$\sigma*\<u\> >_{\KB2,r} \tau$ for some $\tau \in \ET2(x)$.

Let $\rho$ be the next element of $\sigma*\<u\>$ in $\ET2(y)$.
Then
$\ET2(x) = \alpha * \<\rho\> * \beta$ and
$\ET2(y) = \alpha * \<\sigma*\<u\>,\rho\> * \beta$.
By definition of $\ET2$, $\sigma*\<u\> >_{\KB2,r} \rho$.
Since $\sigma*\<u\>$ is maximal in $\ET(y)$ by Lemma \ref{lemma:insert} (1),
there is no $\alpha \ne \<\ \>$ such that $\sigma*\<u\>*\alpha = \rho$.
Hence $\sigma*\<u\> <_{\KB1,r} \rho$.
Hence $\ET2(x) >_{\KB,r} \ET2(y)$.
$\Box$

The next lemma shows that
induction for decreasing transitive sequences
is reduced to
induction for Erd\"os trees with $>_{\KB,r}$.

\begin{Lemma}\label{lemma:ET-ind}
$\HA \prove \forall r \in U.\Ind(\ET2(\DT(U,>_{R_1+ \ldots+ R_k})_{\<r\>}), >_{\KB,r})$ implies
$\HA \prove \Ind(\DT(U,>_{R_1+ \ldots+ R_k}), >_\Ext)$.
\end{Lemma}

{\em Proof.}
In this proof, for simplicity,
we write $\DT$ for $\DT(U,>_{R_1+ \ldots+ R_k})$
and
we also write $\ET_r$ for $\ET2(\DT(U,>_{R_1+ \ldots+ R_k})_{\<r\>})$.

Assume $\HA \prove \forall r\in U.\Ind(\ET_r, >_{\KB,r})$.
Assume
\begin{eq}1
\forall x\in \DT((\forall y<_\Ext x\in \DT.Fy) \imp Fx).
\end{eq}%
in order to show 
\begin{eq}2
\forall x\in\DT.Fx.
\end{eq}%
Define
$
Gy \equiv \forall z \in \DT(z \ne \<\ \> \imp \ET2(z)=y  \imp Fz).
$
We will show
\begin{eq}3
\forall r \in U.\forall x\in\ET_r((\forall y<_{\KB,r} x\in\ET_r.Gy) \imp Gx).
\end{eq}%
Fix $r \in U$ and $x \in\ET_r$ and
assume 
\begin{eq}4
\forall y<_{\KB,r} x\in\ET_r.Gy
\end{eq}%
in order to show $Gx$.
Fix $x_0 \in \DT$ and assume 
$x_0 \ne \<\ \>$ and $\ET2(x_0)=x$ in order to show $Fx_0$.

We can show
\[
\forall y_0<_\Ext x_0\in\DT.Fy_0
\]
as follows.
Let $r$ be $\First(x_0)$.
Fix $y_0 <_\Ext x_0 \in \DT$.
Then $x_0,y_0 \in \DT_{\<r\>}$.
Let $y=\ET2(y_0)$.
By Lemma \ref{lemma:ET}, $y<_{\KB,r} x$.
By \eqref4 and $y \in \ET_r$, $Gy$. Hence $Fy_0$.

By taking $x$ to be $x_0$ in \eqref1, $Fx_0$.
Hence $\forall x_0 \in \DT(x_0 \ne \<\ \> \imp x=\ET2(x_0) \imp Fx_0)$, namely, $Gx$.
We have shown \eqref3.

By \eqref3 and $\forall r\in U.\Ind(\ET_r,>_{\KB,r})$,
we have 
$\forall r\in U.\forall x\in \ET_r.Gx$.

For any $x \in\DT$ such that $x \ne \<\ \>$, by taking 
$r$ to be $\First(x)$ 
and 
$x$ to be $\ET2(x)$ 
in it
we have $G(\ET2(x))$. Hence $Fx$.
For $x = \<\ \>$, by taking $x$ to be $\<\ \>$ in \eqref1, $F\<\ \>$.
Combining them, we have \eqref2.
$\Box$

The next lemma shows induction holds when we restrict the universe.
\begin{Lemma}\label{lemma:ind-universe}
$\HA \prove \Ind(U,>)$ and $\HA \prove V \subseteq U$ imply
$\HA \prove \Ind(V,>)$.
\end{Lemma}

{\em Proof.}
We will show $\Ind(V,>)$ for $F$, namely,
\begin{eq}1
\forall x\in V((\forall y<x\in V.Fy) \imp Fx) \imp \forall x\in V.Fx.
\end{eq}%
Let $Gx$ be $x \in V \imp Fx$.
By $\Ind(U,>)$ for $G$, we have
\[
\forall x \in U((\forall y<x\in U.Gy) \imp Gx) \imp \forall x \in U.Gx.
\]
By predicate logic, it is equivalent to \eqref1.
$\Box$

The next lemma shows induction is implied from
induction for decreasing sequences.
\begin{Lemma}\label{lemma:DS}
$\HA \prove \Ind(\DS(U,>), >_\Ext)$ implies
$\HA \prove \Ind(U, >)$.
\end{Lemma}

{\em Proof.}
In this proof, for simplicity, we write $\DS$ for $\DS(U,>)$.

Assume
\begin{eq}1
\forall x\in U((\forall y<x\in U.Fy) \imp Fx)
\end{eq}%
in order to show $\forall x\in U.Fx$.

Define
$
Gx \equiv F(\Last(x)).
$
We will show
\begin{eq}2
\forall x\in\DS((\forall z<_\Ext x\in\DS.Gz) \imp Gx).
\end{eq}%
Fix $x\in\DS$ and
assume
\begin{eq}3
\forall z<_\Ext x\in\DS.Gz
\end{eq}%
in order to show $Gx$.

We can show 
\begin{eq}4
\forall y<\Last(x) \in U.Fy
\end{eq}%
as follows.
Assume $y<\Last(x)$ in order to show $Fy$.
Then $x *\<y\> \in \DS$ and $x >_\Ext x*\<y\>$.
By taking $z$ to be $x*\<y\>$ in \eqref3, we have $G(x*\<y\>)$.
By definition of $G$, $F(\Last(x*\<y\>))$. Hence $Fy$.
We have shown \eqref4.

By taking $x$ to be $\Last(x)$ in \eqref1, we have $F(\Last(x))$. Hence $Gx$.
Hence we have shown \eqref2.

By $\Ind(\DS,>_\Ext)$ with \eqref2, we have $\forall x\in\DS.Gx$.
By taking $x$ to be $\<x\>$ in it, we have $G(\<x\>)$.
By definition of $G$, we have $Fx$.
$\Box$

The next lemma shows induction for a power of a relation implies
induction for the relation.
\begin{Lemma}\label{lemma:ind-power}
$\HA \prove \Ind(U,>^n)$ implies $\HA \prove \Ind(U,>)$.
\end{Lemma}

{\em Proof.}
We can assume $n>1$.

Assume
\begin{eq}1
\forall x\in U((\forall y<x\in U.Fy) \imp Fx)
\end{eq}%
in order to show $\forall x\in U.Fx$.

We will show
\begin{eq}2
\forall x\in U((\forall y<^n x \in U.Fy) \imp Fx).
\end{eq}%
Fix $x \in U$ and assume
\begin{eq}3
\forall y<^n x \in U.Fy
\end{eq}%
in order to show $Fx$.

By induction on $m$,
we will show
\begin{eq}4
\forall m \le n.\forall w<^{n-m} x \in U.Fw
\end{eq}%

Case 1. $m=0$.

Assume $w<^n x \in U$.
By taking $y$ to be $w$ in \eqref3, $Fw$.

Case 2. $m>0$.

Assume $w<^{n-m} x \in U$ in order to show $Fw$.
We can show $\forall y<w\in U.Fy$ as follows.
Assume $y<w$. 
Then $y<^{n-(m-1)} x$. By IH for $m-1$, $Fy$.

By taking $x$ to be $w$ in \eqref1, $Fw$.

We have shown \eqref4.
By taking $m$ to be $n$  and $w$ to be $x$ in it, $Fx$.
Hence we have shown \eqref2.
By $\Ind(U,>^n)$, $\forall x\in U.Fx$.
$\Box$

Define
\[
\Trans(U,>_R) \equiv \forall xyz \in U(x >_R y \land y >_R z \imp x >_R z), \\
\Decide(U,>_R) \equiv \forall xy \in U(x>_R y \lor \neg(x>_R y)).
\]

The next theorem states that
if some powers of relations $>_{R_i}$ have induction principle,
$>_{R_i}$ are decidable and their union is transitive,
then the union has induction principle.
This theorem is the same as Theorem 6.1 in \cite{Berardi17a} except $\HA$ and the decidability condition $\Decide(U,>_{R_i})$.
\begin{Th}[Podelski-Rybalchenko Theorem for Induction]\label{th:Podelski}
If $\HA \prove \Ind(U,>_{R_1}^{n_1})$, 
$\HA \prove \Decide(U,>_{R_1})$,
\ldots,
$\HA \prove \Ind(U,>_{R_k}^{n_k})$,
$\HA \prove \Decide(U,>_{R_k})$,
and
$\HA \prove \Trans(U, >_{R_1 + \ldots + R_k})$,
then $\Ind(U, >_{R_1 + \ldots + R_k})$.
\end{Th}

{\em Proof.}
We will discuss in $\HA$.

By Lemma \ref{lemma:ind-power}, $\Ind(U,>_{R_i})$.
Define $>_{R_1'}$ as $>_{R_1}$ and
$>_{R_{i+1}'}$ as $(>_{R_{i+1}})-(>_{R_1'}) - \ldots - (>_{R_i'})$.
Then $(>_{R_1'})$, \ldots, $(>_{R_k'})$ are disjoint and
$\forall xy \in U(x >_{R_1\cup\ldots\cup R_k} y \imp
x >_{R_1'+\ldots+R_k'} y)$ by 
$\Decide(U,>_{R_i})$ for $1 \le i \le k$.
Since $(>_{R_i'}) \subseteq (>_{R_i})$, $\Ind(U, >_{R_i'})$.
For simplicity, from now on we write $>_{R_i}$ for $>_{R_i'}$ in this proof.
We will show $\Ind(U, >_{R_1+\ldots+R_k})$.

From $\Ind(U,>_{R_i})$, we have 
$\Ind(\DT(U,>_{R_i}),>_\Ext)$ for $1 \le i \le k$.
By Lemma \ref{lemma:MS}, we have $\forall r\in U.\Ind(\MS_{\<r\>}, >_\Ext)$.
Apparently
$\forall u\in U.\Ind(U,>_{u,\Left})$.
By 
taking $U$ to be $U$, $T$ to be $\MS_{\<r\>}$, and $>_u$ to be $>_{u,\Left}$ in
Theorem \ref{th:KB} for $>_{\KB1,r}$, we have 
$\forall r\in U.\Ind(\MS_{\<r\>}, >_{\KB1,r})$.
By Theorem \ref{th:KB} for $>_{\KB2,r}$, we have 
$\forall r\in U.\Ind(\MS_{\<r\>}, >_{\KB2,r})$ similarly.
Hence
$\forall r\in U.\Ind(\DS(\MS_{\<r\>}, >_{\KB2,r}),>_\Ext)$.
From Lemma \ref{lemma:ind-universe},
we have
$\forall r\in U.\Ind(\DS(\MS_{\<r\>},>_{\KB2,r})_{\<\<r\>\>},>_\Ext)$.
By 
taking 
$T$ to be $\DS(\MS_{\<r\>},>_{\KB2,r})_{\<\<r\>\>}$,
$U$ to be $\MS_{\<r\>}$,
and 
$(>_u)$ to be $(>_{\KB1,r})$
in
Theorem \ref{th:KB} for $>_{\KB,r}$, 
we have 
$\forall r\in U.\Ind(\DS(\MS_{\<r\>},>_{\KB2,r})_{\<\<r\>\>}, >_{\KB,r})$.
Since $\ET2(\DT(U,>_{R_1+ \ldots+ R_k})_{\<r\>}) \subseteq \DS(\MS_{\<r\>},>_{\KB2,r})_{\<\<r\>\>}$,
by Lemma \ref{lemma:ind-universe},
we have
$\forall r\in U.\Ind(\ET2(\DT(U,>_{R_1+ \ldots+ R_k})_{\<r\>}), >_{\KB,r})$.
By Lemma \ref{lemma:ET-ind},
$\Ind(\DT(U,>_{R_1+ \ldots+ R_k}), >_\Ext)$.
By $\Trans(U, >_{R_1 +  \ldots +  R_k})$,
$\DT(U,>_{R_1+ \ldots+ R_k})$ is $\DS(U,>_{R_1+ \ldots+ R_k})$.
Hence we have
$\Ind(\DS(U,>_{R_1+ \ldots+ R_k}), >_\Ext)$.
By Lemma \ref{lemma:DS}, we have $\Ind(U, >_{R_1+ \ldots+ R_k})$.
$\Box$

\section{Proof Transformation}

This section defines main proof transformation from $\CLJIDHA$ to $\LJIDHA$.
The proof is the same as \cite{Berardi17a} except
Theorem \ref{th:Podelski} and
Lemma \ref{lemma:finite}.
First we will define stage numbers and path relations,
and then define proof transformation using them.

For notational convenience,
we assume a cyclic proof $\Pi$ in this section.
Let the buds in $\Pi$ be $J_{1i} \ (i \in I)$ and
the companions be $J_{2j} \ (j \in K)$.
Assume $f:I \to K$ such that the companion of a bud $J_{1i}$ is $J_{2,f(i)}$.

\subsection{Stage Numbers for Inductive Definitions}

In this subsection, we define and discuss stage transformation.

We introduce a stage number to each inductive atomic formula 
so that
the argument of the formula comes into the inductive predicate at the stage 
of the stage number.
This stage number will decrease by a progressing trace.
A proof in $\LJIDHA$ will be constructed by 
using the induction on stage numbers.

First we give stage transformation of an inductive atomic formula.
We assume a fresh inductive predicate symbol $P'$ for
each inductive predicate symbol $P$
and we call it a {\em stage-number inductive predicate symbol}.
$P'(\Vec t,v)$ means that
the element $\Vec t$ comes into $P$ at the stage $v$.
We transform $P(\Vec t)$ into $\exists vP'(\Vec t,v)$.
We call a variable $v$ a {\em stage number}
of $\Vec t$ when $P'(\Vec t,v)$.
$P(\Vec t)$ and $\exists vP'(\Vec t,v)$ will become equivalent 
by inference rules introduced by the transformation of production rules.
We call $P'(\Vec t,v)$ a {\em stage-number inductive atomic formula}.

Secondly 
we give stage transformation of a production rule.
We transform the production of $P$ (the first rule) in Figure \ref{fig:2}
into the production of $P'$ (the second rule) in Figure \ref{fig:2},
where $v,v_1,\ldots,v_m$ are fresh variables.
\begin{figure*}[t]
{\prooflineskip
\[
\infer{P\Vec t}{Q_1\Vec u_1 & \ldots & Q_n\Vec u_n & P_1\Vec t_1 & \ldots & P_m\Vec t_m}
\\
\infer{P'\Vec tv}{Q_1\Vec u_1 & \ldots & Q_n\Vec u_n & v > v_1 & P'_1\Vec t_1v_1 & \ldots & v > v_m & P'_m\Vec t_mv_m & \N v}
\]
}
\caption{Production Rules}\label{fig:2}
\end{figure*}

Next we give stage transformation of a sequent.
For given fresh variables $\Vec v$,
we transform a sequent $J$ into $J^\circ_{\Vec v}$ defined as follows.
We define $\Gamma^\bullet$ 
as the sequent obtained from $\Gamma$
by replacing $P(\Vec t)$ by $\exists vP'(\Vec t,v)$.
For fresh variables $\Vec v$, 
we define $(\Gamma)^\circ_{\Vec v}$ 
as the sequent obtained from $\Gamma^\bullet$
by replacing the $i$-th element 
of the form $\exists vP'(\Vec t,v)$ in the sequent $\Gamma^\bullet$
by $P'(\Vec t,v_i)$.
We define $(\Gamma \prove \Delta)^\bullet$ by $\Gamma^\bullet \prove \Delta^\bullet$,
and define $(\Gamma \prove \Delta)^\circ_{\Vec v}$ by $(\Gamma)^\circ_{\Vec v} \prove \Delta^\bullet$.

We write $(a_i)_{i \in I}$ for the sequence of elements $a_i$ 
where $i$ varies in $I$.
We extend the notion of proofs by allowing open assumptions.
We write $\Gamma \prove_\CLJIDHA \Delta$ with assumptions $(J_i)_{i \in I}$
when there is a proof with assumptions $(J_i)_{i \in I}$ and
the conclusion $\Gamma \prove \Delta$ in $\CLJIDHA$.

\begin{Def}\rm
In a path $\pi$ in a proof,
we define $\Ineq(\pi)$ as the set of the forms
$v > v'$ and $v = v'$ for any stage numbers $v,v'$
eliminated by every case distinction in $\pi$.
\end{Def}

The next lemma shows a stage number is a number.
\begin{Lemma}\label{lemma:N}
(1) $P'\Vec tv \prove_\CLJIDHA \N v$.

(2) $P'\Vec tv \prove_\LJIDHA \N v$.
\end{Lemma}
{\em Proof.}
(1) and (2) are proved by (Case $P'$) and $(\Ind\ P')$ respectively.
$\Box$

An example for $\Ineq(\pi)$ is as follows.
When the path $\pi$ has the case distinction
\[
\Gamma^\circ, u=t(y), \hat v=v,Q_1u_1(y), \ldots, Q_nu_n(y), 
\\ \qquad
v > v_1, P'_1t_1(y)v_1, 
\ldots, v > v_m, P'_mt_m(y)v_m, 
\\ \qquad
\N v \prove \Delta^\bullet
\]
we take $\hat v=v,v>v_1,\ldots,v>v_m$ for $\Ineq(\pi)$.

The proof of the next proposition gives 
{\em stage transformation}
of a proof into
a proof of the stage transformation of the conclusion of the original proof.
We write $\Pi^\circ$ for the stage transformation of $\Pi$.

\begin{Prop}[Stage Transformation]\label{prop:birthtrans1}
For any fresh variables $\Vec v$, 
$\Gamma \prove_\CLJIDHA \Delta$ with assumptions $(\Gamma_i \prove \Delta_i)_{i \in I}$
without any buds
implies 
$(\Gamma)^\circ_{\Vec v} \prove_\CLJIDHA \Delta^\bullet$
with assumptions $(\Ineq(\pi_i),(\Gamma_i)^\circ_{\Vec v_i} \prove \Delta_i^\bullet)_{i \in I}$
without any buds
for some fresh variables $(\Vec v_i)_{i \in I}$,
where $\pi_i$ is the path from the conclusion to the assumption 
$(\Gamma_i)^\circ_{\Vec v_i} \prove \Delta_i^\bullet$.
\end{Prop}

{\em Proof.}
By induction on the proof.
We will transform
the proof $\Pi$ of $\Gamma \prove_\CLJIDHA \Delta$ with assumptions $(J_i)_{i \in I}$ into
the proof of $\Gamma^\circ_{\Vec v} \prove_\CLJIDHA \Delta^\bullet$ with 
assumptions $(\Ineq(\pi_i),(J_i)^\circ_{\Vec v_i})_{i \in I}$ by
transforming each rule as follows.

Case 1. The rule is not ($P_i$ R), (Case), (Cut), 
(Axiom) with a common inductive atomic formula,
or logical rules with
some of the main formula and the auxiliary formulas being
an inductive atomic formula in the antecedent.

Since the main formula and the auxiliary formulas
are not inductive atomic formulas in the antecedent, 
we can just replace each sequent $J$ by $(J)^\circ_{\Vec v}$.
If the rule is an assumption, since $I=\{1\}$,
we take $\Vec v_1$ to be $\Vec v$.

For example,
\[
\infer[(\lor R_l)]{\Gamma \prove P_1t_1 \lor P_2t_2}{
\Gamma \prove P_1t_1
}
\]
is transformed into
\[
\infer[(\lor R_l)]{(\Gamma)^\circ_{\Vec v} \prove \exists v_1P'_1t_1v_1 \lor \exists v_2P'_2t_2v_2}{
(\Gamma)^\circ_{\Vec v} \prove \exists v_1P'_1t_1v_1
}
\]

Case 2. The rule is (Axiom) with a common inductive atomic formula.

We transform
\[
\infer[(Axiom)]{\Gamma,P(\Vec t) \prove P(\Vec t)}{}
\]
into 
\[
\infer[(\exists R)]{\Gamma^\circ_{\Vec v},P'(\Vec t,v) \prove \exists vP'(\Vec t,v)}{
	\infer[(Axiom)]{\Gamma^\circ_{\Vec v},P'(\Vec t,v) \prove P'(\Vec t,v)}{}
}
\]

Case 3. The rule is (Cut),
or logical rules with
some of the main formula and the auxiliary formulas being
an inductive atomic formula in the antecedent.

We replace each sequent $J$ by $(J)^\circ_{\Vec v}$ and
use
\[
\infer[(\exists L)]{\Gamma^\circ_{\Vec v},\exists vP'(\Vec t,v) \prove \Delta^\bullet}{\Gamma^\circ_{\Vec v},P'(\Vec t,v) \prove \Delta^\bullet}
\]

Note that by IH we can assume each $P'(\Vec t,v)$ has a fresh $v$,
so we can use $(\exists L)$.
For example, we transform
\[
\infer[(\neg R)]{\Gamma \prove \neg Pt}{\Gamma,Pt \prove}
\]
into
\[
\infer[(\neg R)]{\Gamma^\circ_{\Vec v} \prove \neg \exists vP'tv}{
\infer[(\exists L)]{\Gamma^\circ_{\Vec v},\exists vP'tv \prove}{
\Gamma^\circ_{\Vec v},P'tv \prove
}}
\]

Case 4. The rule is $(P\ R)$. 
Assume the production rule of $P$ 
and its stage transformation in Figure \ref{fig:2}.
Let $\Sigma_0$ be the set of all assumptions of this production rule of $P'$
and $t_0$ be $v_1+\ldots+v_n+1$.
By $(P'\ R)$ for this production rule,
\[
\Sigma_0 \prove P'\Vec tv.
\]
Since $P_i\Vec t_iv_i \prove \N v_i$, $(\Sigma_1)^\circ_{\Vec v} \prove \N v_i$
where $\Sigma_1$ is $Q_1\Vec u_1, \ldots, Q_m\Vec u_m,
P_1\Vec t_1, \ldots, P_n\Vec t_n$ and $\Vec v$ is $v_1\ldots v_n$.
By (Subst) with $v:=t_0$,
and
(Cut) with $\prove t_0 > v_i$ and $(\Sigma_1)^\circ_{\Vec v} \prove \N t_0$,
\[
(\Sigma_1)^\circ_{\Vec v} \prove P'\Vec tt_0.
\]
By $(\exists R)$, then by $(\exists L)$,
$
(\Sigma_1)^\bullet \prove \exists vP'\Vec tv.
$

By IH,
\[
\Gamma^\circ_{\Vec v} \prove Q_j\Vec u_j, \\
\Gamma^\circ_{\Vec v} \prove \exists v_iP'_i\Vec t_iv_i.
\]
By (Cut),
$
\Gamma^\circ_{\Vec v} \prove \exists vP'\Vec tv.
$

Case 5. The rule is $(\Case\ P)$.
Assume the production rule of $P$ 
and its stage transformation in Figure \ref{fig:2}.
Let $\Vec v$ be $\Vec v'\hat v$.
Let $v_1,\ldots,v_m$ be fresh variables.
Let the rule be
\[
\infer[(\Case\ P)]{\Gamma,P\Vec u \prove \Delta}{
\hbox{case distinctions}
}
\]
with the case distinctions
\[
\Gamma, \Vec u=\Vec t(\Vec y),Q_1\Vec u_1(\Vec y),\ldots, Q_n\Vec u_n(\Vec y), 
\\ \qquad
P_1\Vec t_1(\Vec y), \ldots, P_m\Vec t_m(\Vec y)
\prove \Delta.
\]

Let
\[
\Sigma \equiv (\Vec u=\Vec t(\Vec y), Q_1\Vec u_1(\Vec y), \ldots, Q_n\Vec u_n(\Vec y),
\\ \qquad
P'_1\Vec t_1(\Vec y)v_1, \ldots, P'_m\Vec t_m(\Vec y)v_m), \\
\Sigma_1 \equiv (\hat v=v,v > v_1,\ldots,v > v_m).
\]

By IH with $\Vec v'v_1\ldots v_m$ for the case distinction
we obtain a proof of
\[
\Gamma^\circ_{\Vec v'}, \Sigma \prove \Delta^\bullet
\]
with assumptions $(\Ineq(\pi'_i),(J_i)^\circ_{\Vec v_i})_{i \in I}$
for some $(\Vec v_i)_{i \in I}$
where 
$\pi'_i$ is the path from the transformation of the case distinction
to $(J_i)^\circ_{\Vec v_i}$.
Add $\Sigma_1$
to the antecedent of every sequent in the path $\pi'_i$,
by renaming fresh variables to keep freshness of all fresh variables
in the path.
Then we have a proof of
\[
\Sigma_1,
\Gamma^\circ_{\Vec v'}, \Sigma \prove \Delta^\bullet
\]
with assumptions $(\Sigma_1,\Ineq(\pi_i'),(J_i)^\circ_{\Vec v_i})_{i \in I}$.
We have
\[
\infer[(\Case\ P')]{\Gamma^\circ_{\Vec v}, P'\Vec u\hat v \prove \Delta^\bullet}{
\hbox{case distinctions}
}
\]
with the case distinctions
\[
\Gamma^\circ_{\Vec v}, \Sigma_1,\Sigma, \N v \prove \Delta^\bullet.
\]
By (Wk) with $\N v$ we obtain the case distinction.
By (Case $P'$),
\[
\Gamma^\circ_{\Vec v'}, P'\Vec u\hat v \prove \Delta^\bullet.
\]
Let $\pi_i$ be the path from $\Gamma^\circ, P'\Vec u\hat v \prove \Delta^\bullet$
to $(J_i)^\circ_{\Vec v_i}$.
Then
\[
\Ineq(\pi_i) = (\Sigma_1,\Ineq(\pi'_i)).
\]
Hence we have a proof of
\[
\Gamma^\circ_{\Vec v'}, P'\Vec u\hat v \prove \Delta^\bullet
\]
with assumptions $(\Ineq(\pi_i),(J_i)^\circ_{\Vec v_i})_{i \in I}$.
$\Box$

\subsection{Path Relation}

In this section, we will introduce path relations and discuss them.

We assume a subproof $\Pi_j$ of $\Pi$ 
such that it does not have buds,
its conclusion is $J_{2j}$ and its assumptions are
$J_{1i} \ (i \in I_j)$. 

For $J$ in $\Pi^\circ_j$, we define $\Tilde J$ as $\< v_1,\ldots,v_k \>$ where
$J$ is $\Gamma^\circ_{v_1\ldots v_k} \prove \Delta^\bullet$.

For a path $\pi$ from the conclusion to an assumption in $\Pi_j^\circ$,
we write $\check\pi$ for the corresponding path in $\Pi$.
We extend this notation to a finite composition of $\pi$'s.
By the correspondence $(\check{\ })$,
a stage-number inductive atomic formula in $\Pi^\circ_j$
corresponds to an inductive atomic formula in $\Pi$,
and
a path, a trace, and a progressing trace in $\Pi^\circ_j$
correspond to the same kind of objects in $\Pi$.

\begin{Def}\rm
For a finite composition $\pi$ of paths in $\{ \Pi_j^\circ \ |\ j \in K\}$ such that
$\check\pi$ is a path in the infinite unfolding of $\Pi$,
we define the {\em path relation} $\tilde>_{\pi}$ by
\[
x \tilde>_{\pi} y \equiv
|x|=|\Tilde J_2| \land |y|=|\Tilde J_1| \land
\Land_{F(q_1,q_2)} (x)_{q_2} > (y)_{q_1}
\land
\Land_{G(q_1,q_2)} (x)_{q_2} = (y)_{q_1}
\]
where 
$J_1$ and $J_2$ are the top and bottom sequents of $\pi$ respectively,
$\check J_1$ and $\check J_2$ are those of the path $\check\pi$,
$F(q_1,q_2)$ is that
there is some progressing trace from the $q_2$-th atomic formula in $\check J_2$
to the $q_1$-th atomic formula in $\check J_1$,
$G(q_1,q_2)$ is that
there is some non-progressing trace from the $q_2$-th atomic formula in $\check J_2$ to the $q_1$-th atomic formula in $\check J_1$.
\end{Def}

We define $B_1$ as the set of paths from conclusions to assumptions in 
$\Pi^\circ_j \ (j \in K)$.
We define $B$ as the set of finite compositions of elements in $B_1$ such that
if $\pi \in B$ then $\check\pi$ is a path in the infinite unfolding of $\Pi$.

\begin{Def}\rm\label{def:>}
For $\pi \in B$,
define $x >_{\pi} y$ by
\[
x >_{\pi} y \equiv
(x)_0=j \land (y)_0=f(i) \land  (x)_1 \tilde>_{\pi} (y)_1,
\]
where 
$J_{1i}$ is the top sequent of $\check\pi$, and
$J_{2j}$ is the bottom sequent of $\check\pi$.
\end{Def}

Note that $(\ )_0$ and $(\ )_1$ are operations for a number that represents
a sequence of numbers defined in Section 3.
The first element is a companion number.

\begin{Lemma}\label{lemma:finite}
$\{ >_{\pi} \ |\ \pi \in B \}$ is finite.
\end{Lemma}

{\em Proof.}
Define $C_n$ as $\{ >_{\pi_1\ldots \pi_m} \ |\ m \le n, \pi_i \in B_1 \}$.
Since $>_{\pi}$ is a relation on $N \times N^{\le p}$ where
$p$ is the maximum number of inductive atomic formulas in the antecedents of $\Pi$,
there is $L$ such that $|C_n| \le L$ for all $n$.
Then we have the least $n$ such that $C_{n+1}=C_n$.
Then $|\{ >_{\pi} \ |\ \pi \in B \}|=|C_n|$.
$\Box$

The next proposition shows a sequent is implied from
its stage-number transformation.
\begin{Prop}\label{prop:birthtrans2}
For any fresh variables $\Vec v$,
$\Gamma^\circ_{\Vec v} \prove_\LJIDHA \Delta^\bullet$ 
implies $\Gamma \prove_\LJIDHA \Delta$.
\end{Prop}

{\em Proof.}
First we can show $\Gamma^\bullet \prove_\LJIDHA \Delta^\bullet$
by 
\[
\infer[(\exists L)]{
\exists v P'\Vec tv,\Gamma^\bullet \prove \Delta^\bullet}{
P'\Vec tv,\Gamma^\bullet \prove \Delta^\bullet
}
\]

Secondly
by
\[
\prove_\LJIDHA P\Vec t \lequiv \exists vP'(\Vec t,v)
\]
we have 
\[
\Gamma \prove_\LJIDHA \Delta.
\]
$\Box$

The next lemma shows $\tilde>_\pi$ is an abstraction of $\Ineq(\pi)$.
\begin{Lemma}\label{lemma:decrease}
For a proof $\Pi$ without any buds,
if $\pi$ is a path from $(J_2)^\circ_{\Vec x}$ to $(J_1)^\circ_{\Vec y}$ in $\Pi^\circ$,
\[
\Ineq(\pi) \prove_\HA \<\Vec x\>  \tilde>_{ \pi} \<\Vec y\>.
\]
\end{Lemma}

{\em Proof.}
By induction on $|\pi|$. 

Case 1. $|\pi|=0$.
$J_1 = J_2$. $x \tilde>_{ \pi} y$ is
$|x|=m \land |y|=m \land (x)_0=(y)_0 \land \ldots \land (x)_{m-1}=(y)_{m-1}$
where $m$ is the number of stage-number inductive atomic formulas in $(J_1)^\circ_{\Vec y}$.
Hence $\prove_\HA x \tilde>_{ \pi} y$.

Case 2. $|\pi|>0$.
Consider cases according to the last rule of $\pi$.
Let $\pi=\pi_1\pi_2$ such that $|\pi_1|=1$.
Let the top sequent of $\pi_1$ be $(J_3)^\circ_{\Vec z}$.
Let $x$,$y$,$z$ be $\<\Vec x\>$, $\<\Vec y\>$, $\<\Vec z\>$ respectively.

By IH, $\Ineq(\pi_2) \prove_\HA z \tilde>_{ \pi_2} y$.

We will show $\Ineq(\pi_1) \prove_\HA x \tilde>_{ \pi_1} z$.
Since the rule that changes the stage number is only (Case),
we will show only the case (Case).
Assume the production rule of $P$ 
and its stage transformation in Figure \ref{fig:2}.
Let the path for the rule (Case $P'$) be
\[
\infer{\Gamma^\circ_{\Vec v}, P'\Vec u\hat v \prove \Delta^\bullet}{
\Gamma^\circ_{\Vec v}, \Sigma_1,\Sigma, \N v \prove \Delta^\bullet
}
\]
where
\[
\Sigma \equiv (\Vec u=\Vec t(\Vec y), Q_1\Vec u_1(\Vec y), \ldots, Q_n\Vec u_n(\Vec y),
\\ \qquad
P'_1\Vec t_1(\Vec y)v_1, \ldots, P'_m\Vec t_m(\Vec y)v_m), \\
\Sigma_1 \equiv (\hat v=v,v > v_1,\ldots,v > v_m).
\]
Then $x=\<\Vec v,\hat v\>$ and $z=\<\Vec v,v_1,\ldots,v_m\>$, and
\[
x \tilde>_{ \pi_1} z \lequiv v_1 < \hat v \land \ldots \land v_m < \hat v.
\]
Since $\Ineq(\pi_1) \equiv \Sigma_1$,
$
\Ineq(\pi_1) \prove_\HA x \tilde>_{ \pi_1} z.
$
Since $\Ineq(\pi) = (\Ineq(\pi_1), \Ineq(\pi_2))$,
$
\Ineq(\pi) \prove_\HA 
x \tilde>_{ \pi_1} z \tilde>_{ \pi_2} y.
$
Since 
$(\tilde>_{ \pi_1}) \circ (\tilde>_{ \pi_2}) \subseteq (\tilde>_{\pi_1\pi_2})$,
$
\Ineq(\pi) \prove_\HA x \tilde>_{ \pi_1\pi_2} y.
$
$\Box$

The next lemma is the only lemma that uses the global trace condition.

\begin{Lemma}\label{lemma:globaltrace}
For all $\pi \in B$, there is $n>0$ such that
$\prove_\HA \Ind(U,>_\pi^n)$.
\end{Lemma}

{\em Proof.}
Let the bottom sequent of $\pi$ be $J_{2j}$ and the top sequent be $J_{1i}$.
Let the companion of $J_{1i}$ be $J_{2k}$.

Case 1. $j \ne k$.

Assume
\[
H \equiv \forall x \in U.(\forall y <_\pi x \in U.Fy) \imp Fx
\]
and fix $x \in U$.

Assume $y <_\pi x \in U$.
By taking $x$ to be $y$ in $H$, 
\[
(\forall z <_\pi y \in U.Fz) \imp Fy.
\]
By $\neg (z <_\pi y)$ from $y <_\pi x$ and $j \ne k$, we have $Fy$.
Hence $H \prove_\HA \forall y <_\pi x \in U.Fy$.

By taking $x$ to be $x$ in $H$, we have $Fx$.
Hence $H \prove_\HA \forall x \in U.Fx$.
Hence $\prove_\HA \Ind(U,>_\pi,F)$.
We can take $n$ to be 1.

Case 2. $j = k$.

By applying the global trace condition to the infinite path $\check\pi^\omega$,
there is a progressing trace in the path.
Hence there are $n,m,q$ such that
the trace passes
the $q$-th stage-number inductive atomic formula in the top sequent of $\pi^m$
and
the $q$-th stage-number inductive atomic formula in the top sequent of $\pi^{m+n}$.

Define $x <_q y$ by $((x)_1)_q < ((y)_1)_q$.
By mathematical induction we can easily show
\[
(\forall x\in U.(\forall y <_q x \in U.Fy) \imp Fx) \imp \forall x\in U.Fx.
\]
If $y <_\pi^n x$, then
$y <_{\pi^n} x$, and hence
$((y)_1)_q < ((x)_1)_q$, which implies $y <_q x$. Therefore
\[
(\forall x\in U.(\forall y <_\pi^n x \in U.Fy) \imp Fx) \imp \forall x\in U.Fx,
\]
which is $\Ind(U,>^n_\pi,F)$.
$\Box$

We define $>_\Pi$ as $\bigcup \{ >_{\pi} \ |\ \pi \in B \}$.
Note that $>_\Pi$ is transitive,
since 
the top sequent of $\pi_1$
is the bottom sequent of $\pi_2$ by the first element, and
$((>_{\pi_1}) \circ (>_{\pi_2})) \subseteq (>_{\pi_1\pi_2})$.

\subsection{Proof Transformation}

This section gives main proof transformation.

The next lemma shows
we can replace (Case) rules of $\CLJIDHA$ 
by (Ind) rules of $\LJIDHA$.
\begin{Lemma}\label{lemma:case-ind}
If there is a proof
with some assumptions
and without any buds in $\CLJIDHA$,
then
there is a proof of the same conclusions with the same assumptions
in $\LJIDHA$.
\end{Lemma}

{\em Proof.}
It is sufficient to replace the rule (Case) by the rule $(\Ind)$.
This is straightforward and
has been proved by Lemma 4.1.4 in \cite{Brotherston-phd}.
We give only a proof idea here.

Assume the production rule of $P$ in Figure \ref{fig:2}.
We can replace
{
\[
\infer[(Case)]{\Gamma,P\Vec u \prove \Delta}{
	\infer*[\Pi_i]{\Gamma,\Vec u=\Vec t(\Vec y),\Sigma \prove \Delta}{}
	&
	(\forall i \in I)
}
\]
}
by
{
\[
\infer[(\imp L)(\forall L)(Cut)]{\Gamma,P\Vec u \prove \Delta}{
\infer[(\Ind\ P)]{P\Vec u \prove F\Vec u}{
\infer*[\Pi'_i]{
\Sigma' \prove F\Vec t(\Vec y)}{}
&
(\forall i \in I)
}}
\]
}
where
\[
F'\Vec z \equiv \forall \Vec x(\Vec z=\Vec u \imp \Gamma \imp \Delta), \\
F\Vec z \equiv P\Vec z \land F'\Vec z.
\]
$\Box$

The next is the main lemma and shows each bud in a cyclic proof
is provable in $\LJIDHA$.
\begin{Lemma}\label{lemma:prooftrans}
For every bud $J$ of a proof in $\CLJIDHA$ and fresh variables $\Vec v$,
$(J)^\circ_{\Vec v}$ is provable in $\LJIDHA$.
\end{Lemma}

{\em Proof.}
For $j \in K$, define
\[
G_jx \equiv \forall \Vec z(x=\<\Vec v\> \imp (\Gamma_{2j})^\circ_{\Vec v} \imp \Delta_{2j}^\bullet)
\]
where $\Vec z$ is $\FV((\Gamma_{2j})^\circ_{\Vec v},\Delta_{2j}^\bullet)$.

Define
\[
Gx_0x \equiv \Land_{j \in K} (x_0=j \imp G_jx), \\
Hx_0x \equiv \forall y_0y.
\<y_0,y\> <_\Pi \<x_0,x\> \imp Gy_0y.
\]

We will show that 
for every companion $J$ in $\Pi$ and fresh variables $\Vec v$,
$(J)^\circ_{\Vec v}$ is provable in $\LJIDHA$.
Fix a companion in $\Pi$ and fresh variables $\Vec v$.
Let the companion be $J_{2j}$.
We will construct a proof of $(J_{2j})^\circ_{\Vec v}$ in $\LJIDHA$.
We have a subproof $\Pi_j$ of $\Pi$ 
such that it does not have buds,
its conclusion is $J_{2j}$ and its assumptions are
$J_{1i} \ (i \in I_j)$. 
By Proposition \ref{prop:birthtrans1},
\[
\infer*[\Pi_j^\circ]{(\Gamma_{2j})^\circ_{\Vec v} \prove \Delta_{2j}^\bullet}{
(\Ineq(\pi_{ji}), (\Gamma_{1i})^\circ_{\Vec v_i} \prove \Delta_{1i}^\bullet)
_{i \in I_j}
}
\]
for some fresh variables $(\Vec v_i)_{i \in I_j}$,
where
$\pi_{ji}$ is the path from $J_{2j}$ to $J_{1i}$ in $\Pi_j$.
Next we transform $\Pi_j^\circ$ 
into the ($\LJIDHA$)-proof $\Pi'_j$ 
with the same conclusion and the same assumptions
by Lemma \ref{lemma:case-ind}.

Next we add
\[
x_0=j,x=\<\Vec v\>,Hx_0x
\]
to every antecedent in $\Pi'_j$ 
to obtain 
\[
\infer*[\Pi''_j]{x_0=j,x=\<\Vec v\>,Hx_0x,(\Gamma_{2j})^\circ_{\Vec v} \prove \Delta^\bullet_{2j}}{
(x_0=j,x=\<\Vec v\>,Hx_0x,\Ineq(\pi_{ji}),(\Gamma_{1i})^\circ_{\Vec v_i} \prove \Delta^\bullet_{1i})
_{i \in I_j}
}
\]

Let $\Pi_{ij}$ be the proof in Figure \ref{fig:3}.
\begin{figure*}[t]
{\tiny
\[
\infer[(\imp L)(\forall L)(Cut)]{x_0=j,x=\<\Vec v\>,Hx_0x,\Ineq(\pi_{ji}),(\Gamma_{1i})^\circ_{\Vec v_i} \prove \Delta^\bullet_{1i}}{
\infer[(Cut)(Subst)]{x_0=j,x=\<\Vec v\>,Hx_0x,\Ineq(\pi_{ji}) \prove Gf(i)\<\Vec v_i\>}{
	\infer[(Wk)(=L)]{x=\<\Vec v\>,y=\<\Vec v_i\>,\Ineq(\pi_{ji}) \prove x \tilde>_{\pi_{ji}} y}{\hbox{Lemma \ref{lemma:decrease}}}
	&
	\infer[(\imp L)(\forall L)(Cut)]{x_0=j,y_0=f(i),
	\<x_0,x\> >_{\pi_{ji}} \<y_0,y\>, Hx_0x \prove Gy_0y}{}
}}
\]
}
\caption{Subproof $\Pi_{ij}$}\label{fig:3}
\end{figure*}
We have a proof $\Pi'''$ of $\forall x_0x.Gx_0x$ 
with the assumption $\Ind(>_\Pi,G)$
in Figure \ref{fig:1}.
\begin{figure*}[t]
\[
\infer[(\imp R)(\forall R)(Cut)]{\prove \forall x_0x.Gx_0x}{
	\prove \Ind(>_\Pi,G)
	&
	\infer[(\land R)]{Hx_0x \prove Gx_0x}{
		(\forall j \in K)
		&
		\infer[(\imp R)(\forall R)(\imp R)(\imp R)]{Hx_0x \prove x_0=j \imp G_jx}{
		\infer*[\Pi''_j]{x_0=j,x=\<\Vec v\>,Hx_0x,(\Gamma_{2j})^\circ_{\Vec v} \prove \Delta^\bullet_{2j}}{
		\left(
		\infer*[\Pi_{ij}]{x_0=j,x=\<\Vec v\>,Hx_0x,\Ineq(\pi_{ji}),(\Gamma_{1i})^\circ_{\Vec v_i} \prove \Delta^\bullet_{1i}}{}
		\right)
		_{i \in I_j}
}}}}
\]
\caption{Proof $\Pi'''$}\label{fig:1}
\end{figure*}

By applying Theorem \ref{th:Podelski} to
$U = \N \times \N^{\le p}$ and
$\{ >_\pi \ |\ \pi \in B \}$, 
\[
\HA \prove \Ind(U,>_\pi^{n_\pi}) \ (\forall \pi \in B) 
\]
implies
\[
\HA \prove \Ind(U, >_\Pi).
\]
By Lemma \ref{lemma:globaltrace}, we have
\[
\prove_\HA \Ind(U,>_\pi^{n_\pi}) \  (\forall \pi \in B).
\]
By the definition of $>_\pi$,
\[
\prove_\HA \Decide(U, >_\pi)
\]
for all $\pi \in B$.

Then
\[
\infer[(\imp L)(\forall L)(Cut)]{(J_{2j})^\circ_{\Vec v}}{
\infer[(\imp L)(\land L)(Cut)]{\prove G_j\<\Vec v\>}{
\infer[(\forall L)(Cut)]{\prove Gj\<\Vec v\>}{
\infer*[\Pi''']{\prove \forall x_0x.Gx_0x}{
\infer{\prove \Ind(U,>_\Pi,G)}{
	\hbox{Lemma \ref{lemma:globaltrace}}
	&
	\prove \Decide(U, >_\pi)
	&
	\hbox{Theorem \ref{th:Podelski}}
}}}}}
\]

We have shown that 
for every companion $J$ in $\Pi$ and fresh variables $\Vec v$,
$(J)^\circ_{\Vec v}$ is provable in $\LJIDHA$.

Fix a bud be $J_{1k}$ in $\Pi$ and fresh variables $\Vec v$.
Let $J_{2j}$ be the companion of the bud.
Since $(J_{2j})^\circ_{\Vec v}$ is provable, 
$(J_{1k})^\circ_{\Vec v}$ is provable in $\LJIDHA$.
$\Box$

We write $\LJIDHA+(\Sigma,\Phi)$ for the system
$\LJIDHA$ with the signature $\Sigma$ and the set $\Phi$ of production rules.
Similarly we write $\CLJIDHA+(\Sigma,\Phi)$.
For simplicity, in $\Phi$ we write only $P$ for the set of production rules
for $P$.
We define
$\Sigma_N = \{0,s,+,\times,<,N\}$ and
$ \Phi_N = \{N\}.$
We write $P''$ for $(P')'$.

The next is the main proposition stating a cyclic proof is transformed
into an $(\LJIDHA)$-proof with stage-number inductive predicates.
\begin{Prop}\label{prop:equiv}
If a sequent $J$ is provable in
$\CLJIDHA+(\Sigma_N \cup \{ \Vec P \}, \Phi_N \cup \{ \Vec P \})$,
then
$J$ is provable in 
$\LJIDHA+(\Sigma_N \cup \{N', \Vec P, \Vec P' \}, \Phi_N \cup \{ \Vec P, \Vec P' \})$
where $N', \Vec P'$ are the stage-number inductive predicates 
of $N, \Vec P$.
\end{Prop}

{\em Proof.}
Let $\Pi_1$ be the cyclic proof of $\Gamma \prove_\CLJIDHA \Delta$.
Let $(J_{1i})_{i \in I}$ be all the buds in $\Pi_1$.
Define $\Pi_2$ be a proof obtained from $\Pi_1$ by removing 
all bud-companion relations.
Then
$\Pi_2$ is a proof of $\Gamma \prove \Delta$
with assumptions $(J_{1i})_{i \in I}$ and without buds in $\CLJIDHA$.
Choose fresh variables $\Vec v$.
By Proposition \ref{prop:birthtrans1}, there is $(\Vec v_i)_{i \in I}$ such that
$\Gamma^\circ_{\Vec v} \prove \Delta^\bullet$
with assumptions $(\Ineq(\pi_i),(J_{1i})^\circ_{\Vec v_i})_{i \in I}$ and without buds in $\CLJIDHA$.
By Lemma \ref{lemma:case-ind}, 
$\Gamma^\circ_{\Vec v} \prove \Delta^\bullet$
with assumptions $(\Ineq(\pi_i),(J_{1i})^\circ_{\Vec v_i})_{i \in I}$ and without buds is provable in $\LJIDHA$.
By (Wk),
there is a proof $\Pi_3$ of $\Gamma^\circ_{\Vec v} \prove \Delta^\bullet$
with assumptions $((J_{1i})^\circ_{\Vec v_i})_{i \in I}$ in $\LJIDHA$.
By Lemma \ref{lemma:prooftrans},
$(J_{1i})^\circ_{\Vec v_i}$ is provable in $\LJIDHA$.
Combining them with $\Pi_3$,
$\Gamma^\circ_{\Vec v} \prove \Delta^\bullet$ is provable in $\LJIDHA$.
By Proposition \ref{prop:birthtrans2},
$\Gamma \prove \Delta$ is provable in $\LJIDHA$.
$\Box$

The next shows conservativity for 
stage-number inductive predicates.
\begin{Prop}[Conservativity of $N'$ and $P''$]\label{prop:conservative}
Let 
$
\Sigma = \Sigma_N \cup \{\Vec Q,\Vec P, \Vec P'\}$,
$\Phi = \Phi_N \cup \{ \Vec P, \Vec P'\}$,
$\Sigma' = \Sigma \cup \{N',\Vec P''\}$, and
$\Phi' = \Phi \cup \{N', \Vec P''\}$.
Then
$\LJIDHA+(\Sigma',\Phi')$
is conservative over
$\LJIDHA+(\Sigma,\Phi)$.
\end{Prop}

{\em Proof.}
Define $\Tilde{(\ )}$ by replacing
$N'$ by $\lambda xy.Nx \land Ny \land x \le y$ and
replacing $P_j''$ by $\lambda\Vec xyz.P_j'\Vec xy \land Nz \land y<z$
for all $j$.

By induction on a proof,
we will show that
$\LJIDHA+(\Sigma',\Phi') \prove J$ implies
$\LJIDHA+(\Sigma,\Phi) \prove \Tilde J$.

Use case analysis by the last rule.
If the last rule is not $(\Ind\ N')$ or $(\Ind\ P_j'')$, the claim follows immediately from IH.

Case 1. The last rule is $(\Ind\ N')$.

Let the rule be
\[
\infer[(\Ind\ N')]{\Gamma,N'ab \prove Fab}{
	\Gamma,Nv \prove F0v
	&
	\Gamma,Fxv_1,v_1<v,Nv \prove Fsxv
}
\]
By IH for these two premises, we have
\begin{eq}1
\Tilde \Gamma,Nv \prove \Tilde F0v
\end{eq}%
and
\begin{eq}2
\Tilde \Gamma,\Tilde Fxv_1,v_1<v,Nv \prove \Tilde Fsxv.
\end{eq}%

Define $F'a$ be $\forall z\in N.(a \le z \imp \Tilde Faz)$.
By $(\Ind\ N)$ we will show $\Tilde\Gamma, Na \prove F'a$.

We can show the first premise $\Tilde\Gamma \prove F'0$,
since it immediately follow from \eqref1.

We can show the second premise $\Tilde\Gamma,Nx,F'x \prove F'sx$ as follows.
Assume $\Tilde\Gamma,Nx,F'x$ and
fix $z \in N$ and assume $sx \le z$ in order to show $\Tilde Fsxz$.
Then there is $z' \in N$ such that $z=sz'$.
Then $x \le z'$.
By $F'x$, $\Tilde Fxz'$.
By taking $v_1$ to be $z'$ and $v$ to be $z$ in \eqref2, $\Tilde Fsxz$.

By $(\Ind\ N)$, we have $\Tilde\Gamma, Na \prove F'a$.
Hence 
$\Tilde\Gamma, Na, Nb, a \le b \prove \Tilde Fab$.
Hence
$\Tilde{(\Gamma, N'ab)} \prove \Tilde{Fab}$.

Case 2. The last rule is $(\Ind\ P_j'')$.

Let the rule be
\[
\infer[(\Ind\ P_j'')]{\Gamma, P_j''\Vec abc \prove F_j\Vec abc}{
J_i 
& 
(\forall i \in I)
}
\]

Define $F_k'\Vec ab$ be $Nb \land \forall z \in N(b<z \imp \Tilde F_k\Vec abz)$
for all $k$.

We will show
$\Tilde\Gamma,P_j'\Vec ab \prove F'_j\Vec ab$
by
\[
\infer[(\Ind\ P_j')]{\Tilde\Gamma,P_j'\Vec ab \prove F'_j\Vec ab}{
J'_i
&
(\forall i \in I)
}
\]
We will show each $J'_i$.
Let $J'_i$ be
\[
\Tilde\Gamma, Q_1\Vec u_1, \ldots,Q_n\Vec u_n,
F_1'\Vec t_1v_1, v_1<v, \ldots, F_m'\Vec t_mv_m, v_m<v, Nv \prove F'\Vec tv.
\]
and
the production rules be
{\prooflineskip
\[
\infer{P\Vec t}{Q_1\Vec u_1 & \ldots & Q_n\Vec u_n &
	P_1\Vec t_1 & \ldots & P_m\Vec t_m
}
\\
\infer{P'\Vec tv}{Q_1\Vec u_1 & \ldots & Q_n\Vec u_n &
	P_1'\Vec t_1v_1 & v_1<v & \ldots & P_m'\Vec t_mv_m & v_m<v & Nv
}
\\
\infer{P''\Vec tvw}{
\parbox{10cm}{$Q_1\Vec u_1 \quad \ldots \quad Q_n\Vec u_n \quad
	P_1''\Vec t_1v_1w_1 \quad w_1 < w \quad v_1<v \quad \ldots \quad 
\hfil\break
	P_m''\Vec t_mv_mw_m \quad w_m < w \quad v_m<v \quad N'vw_0 \quad w_0<w \quad Nw$}
}
\]
}

We can assume the production rule for $J'_i$ 
is the stage-number transformation of the production rule for $J_i$.
Then $J_i$ is
\[
\Gamma, Q_1\Vec u_1, \ldots,Q_n\Vec u_n,
	F_1\Vec t_1v_1w_1, w_1 < w, v_1<v, \ldots,
\\ \qquad
	F_m\Vec t_mv_mw_m, w_m < w, v_m<v, N'vw_0, w_0<w, Nw
\prove F\Vec tvw
\]
By IH for $J_i$ we have $\Tilde{J_i}$, namely
\[
\Tilde\Gamma, Q_1\Vec u_1, \ldots,Q_n\Vec u_n,
\\
	\Tilde F_1\Vec t_1v_1w_1, w_1 < w, v_1<v, \ldots,
\Tilde F_m\Vec t_mv_mw_m, w_m < w, v_m<v, 
\]
\begin{eq}4
Nv, Nw_0, v\le w_0, w_0<w, Nw
\prove \Tilde F\Vec tvw.
\end{eq}%

We can show $J'_i$ as follows.
Assume the antecedent,
fix $z$, and assume $Nz$ and $v<z$ in order to show $\Tilde F\Vec tvz$.
Take $w$ to be $z$, $w_l$ to be $sv_l$, and $w_0$ to be $v$ in \eqref4.
Then 
$w_l<w$ by $v_l<v$ and $v<z$.
We also have
$w_0<w$ by $v<z$.
Moreover
$\Tilde F_l\Vec t_lv_lw_l$ by $F_l'\Vec t_lv_l$.
Hence $\Tilde F_l\Vec tvz$ by \eqref4.
Hence we have shown $J'_i$.

By $(\Ind\ P')$, we have $\Tilde\Gamma,P_j'\Vec ab \prove F_j'\Vec ab$.
Hence $\Tilde\Gamma,P_j'\Vec ab,Nc,b<c \prove \Tilde F_j\Vec abc$.
Hence we have $\Tilde{(\Gamma, P_j''\Vec abc)} \prove \Tilde{F_j\Vec abc}$.
$\Box$

The next main theorem shows Brotherston-Simpson conjecture under arithmetic for intuitionistic logic.
\begin{Th}[Equivalence of $\LJIDHA$ and $\CLJIDHA$]\label{th:equiv}
Let
$\Sigma = \Sigma_N \cup \{\Vec Q,\Vec P, \Vec P'\}$ and
$\Phi = \Phi_N \cup \{ \Vec P, \Vec P'\}$.
Then
the provability of $\CLJIDHA+(\Sigma,\Phi)$ is the same as
that of $\LJIDHA+(\Sigma,\Phi)$.
\end{Th}

{\em Proof.}
(1) $\LJIDHA+(\Sigma,\Phi)$ to $\CLJIDHA+(\Sigma,\Phi)$.

For this claim, we can obtain a proof 
from the proof of Lemma 7.5 in \cite{Brotherston11a}
by restricting every sequent into intuitionistic sequents
and replacing
$\LKID+(\Sigma,\Phi)$ and $\CLKIDomega+(\Sigma,\Phi)$ by
$\LJID+(\Sigma,\Phi)$ and $\CLJIDomega+(\Sigma,\Phi)$ respectively.

(2) $\CLJIDHA+(\Sigma,\Phi)$ to $\LJIDHA+(\Sigma,\Phi)$.

Let
$\Sigma' = \Sigma \cup \{N',\Vec P''\}$ and
$\Phi' = \Phi \cup \{N', \Vec P''\}$.
Assume $J$ is provable in $\CLJIDHA+(\Sigma,\Phi)$.
By Proposition \ref{prop:equiv},
$J$ is provable in $\LJIDHA+(\Sigma',\Phi')$.
By Proposition \ref{prop:conservative},
$J$ is provable in $\LJIDHA+(\Sigma,\Phi)$.
$\Box$

\section{Conclusion}

We have studied Brotherston-Simpson conjecture for intuitionistic logic.
We have pointed out that the countermodel of \cite{Berardi17}
shows the Brotherston-Simpson conjecture for intuitionistic logic is false in general.
We have shown $\HA$-provability of
Kleene-Brouwer theorem for induction and
Podelski-Rybalchenko theorem for induction.
By using them,
we have shown the conjecture for intuitionistic logic is true under
arithmetic,
namely, the provability of the intuitionistic
cyclic proof system is the same as that of the
intuitionistic system of Martin-Lof's inductive definitions when
both systems contain Heyting arithmetic.


\end{document}